\documentstyle[epsfig,12pt]{article}

\def\sla{\raise.15ex\hbox{$/$}\kern-.57em}

\def\Box{\kern1pt\vbox{\hrule height 1.2pt\hbox{\vrule width 1.2pt\hskip 3pt
   \vbox{\vskip 6pt}\hskip 3pt\vrule width 0.6pt}\hrule height 0.6pt}\kern1pt}
\def\gtwid{\mathrel{\raise.3ex\hbox{$>$\kern-.75em\lower1ex\hbox{$\sim$}}}}
\def\ltwid{\mathrel{\raise.3ex\hbox{$<$\kern-.75em\lower1ex\hbox{$\sim$}}}}
\def\Box{\kern1pt\vbox{\hrule height 1.2pt\hbox{\vrule width 1.2pt\hskip 3pt
   \vbox{\vskip 6pt}\hskip 3pt\vrule width 0.6pt}\hrule height 0.6pt}\kern1pt}

\def\d{\delta}
\def\d1{\delta^+}
\def\d2{\delta^-}
\def\O{\Omega}

\newcommand\be{\begin{equation}}
\newcommand\ee{\end{equation}}
\def\bea{\begin{eqnarray}}

\begin{document}
\begin{titlepage}
\begin{flushright}
hep-th/0108090 \\ UFIFT-HEP-01-12 \\ CRETE-01-11 
\end{flushright}
\begin{center}
\bf{Pair creation and axial anomaly in light-cone $QED_2$}
\end{center}

\vspace{0.5cm}

\begin{center}
T. N. Tomaras$^{\dagger}$ and N. C. Tsamis$^{\ddagger}$
\end{center}
\begin{center}
\it{Department of Physics and Institute of Plasma Physics \\
University of Crete and FO.R.T.H. \\ 
P.O.Box 2208, 710 03 Heraklion, Crete; GREECE.}
\end{center}

\vspace{0.3cm}

\begin{center}
R. P. Woodard$^*$
\end{center}
\begin{center}
\it{Department of Physics, University of Florida \\ 
Gainesville, FL 32611; UNITED STATES.}
\end{center}

\vspace{0.7cm}

\begin{center}
ABSTRACT
\end{center}
\hspace*{.3cm} The 1+1 dimensional massive Dirac equation is solved exactly 
in light-cone coordinates for $x^+ > 0$ and $x^- > -L$, in the presence of an 
arbitrary $x^+$ dependent electric field. Our solution resolves the ambiguity 
at $p^+ = 0$. We also obtain the one loop rate for pair production for a 
positive electric field, compute the expectation values of the vector and 
axial vector currents, and recover the well known anomaly $e^2 E/\pi$ in the 
divergence of the latter. A final intriguing result is that the theory seems 
to exhibit a phase transition in the limit of infinite $L$.

\vspace{1cm}

\begin{flushleft}
\end{flushleft}
\begin{flushleft}
$^{\dagger}$ e-mail: tomaras@physics.uoc.gr \\
$^{\ddagger}$ e-mail: tsamis@physics.uoc.gr \\
$^*$ e-mail: woodard@phys.ufl.edu
\end{flushleft}
\end{titlepage}

\section{Introduction}

The phenomenon of electron-positron pair production in an external electric 
field has already a long history. What happens initially when a prepared state 
is released in the presence of a homogeneous electric field is that 
electron-positron pairs emerge from the vacuum to form a current which 
diminishes the electric field. If the state is released on a surface of 
constant $x^0$ with no initial charge then the electric field at later times 
depends only upon $x^0$. This process was considered long before the 
ultraviolet problem of quantum electrodynamics was resolved 
\cite{Klein,Sauter,Wolkow}. Schwinger invented what we now know as the in-out 
background field effective action to compute the rate of particle production 
per unit volume in the presence of a strictly constant electric field 
\cite{Schwinger}. Since then a variety of articles 
\cite{Brezin}--\cite{Kluger2} and monographs \cite{Greiner,Fradkin} have 
treated the issue of what happens when the effect becomes strong.

In the light-cone analog to this process a source-free state is released on a 
surface of constant $x^+ \equiv (x^0 + x^1)/\sqrt{2}$ in the presence of 
a homogeneous electric field which is parallel to $x^1$. The resulting 
evolution yields a homogeneous electric field which depends upon $x^+$ rather 
than $x^0$, and has potentially interesting applications in the study of the 
onset and evolution of electric fields in the vicinity of a star formed by the 
collapse of charged matter \cite{damour}. An amazing feature of Dirac theory 
in {\it any} such $x^+$ dependent background is that the mode functions are 
simple \cite{ttw}. This fact had been noted previously for the special case of 
a constant electric field \cite{artru,Indians1,Indians2}, although without 
resolving the ambiguity at $p^+ = 0$. By allowing the electric field to depend
arbitrarily upon $x^+$ it is for the first time possible to study back-reaction
{\it analytically} in the presence of a class of backgrounds which is 
guaranteed to include the actual solution.

A crucial insight in resolving the ambiguity at $p^+ = 0$ was that the 
light-cone evolution problem is not well posed with only initial value data 
from the surface at $x^+ = 0$ \cite{ttw}. One must also provide data on a
surface of constant $x^- \equiv (x^0 - x^1)/\sqrt{2}$, which we took to be
$x^- = -L$. Incidentally, this fact has been noted before in \cite{neville}, 
while an early attempt to analyze its consequences in the quantum theory of 
massless fermions, may be found in \cite{mccartor}. Although correct, our 
previous solution for the electron field operator was only valid in the 
distributional limit of $L \longrightarrow \infty$. This restricted its to 
use to expectation values of nonsingular operators such as $J^+$, but 
prevented us from evaluating operators such as $J^-$ which diverge as $L$ 
goes to infinity. In the present work we remove this restriction by deriving 
an exact operator solution which is valid for all $L$. The result is a vast 
increase in the range of applications which we will exploit in a number of 
ways for the special case of 1+1-dimensional, massive quantum electrodynamics 
($QED_2$). Note, however, that the exact operator solution is valid for any 
spacetime dimension, as well as for zero mass.

This paper contains seven sections of which this introduction is the first. 
Section 2 begins by giving our light-cone coordinate and gauge conventions,
which we have converted from those of our initial treatment \cite{ttw} to 
standard notation. This section also presents the complete operator solution 
for Dirac theory in the presence of an arbitrary $x^+$ dependent electric 
field background, expressed in terms of the field operators on the surfaces of 
$x_+ = 0$ and $x_- = -L$. Section 3 gives the quantum operator algebra and the 
conditions which define the initial, source-free state. We use these results 
in Section 4 to give an explicit, analytic derivation for the probability of 
particle production in our general background. In Section 5 we compute the one 
loop expectation value of the vector current operator and show that it is 
conserved. The fact that our results depend non-analytically upon the
background field in the large $L$ limit seems to indicate that we are seeing
an infinite volume phase transition. In Section 6 we compute the one loop
expectation values of the axial currents and of the pseudoscalar bilinear. We
exploit these results to derive the standard formula for the axial vector 
anomaly for the first time ever on the light-cone with $m \neq 0$.\footnote{
The simpler massless case has already been given a satisfactory treatment 
\cite{McC,NaMc}.} Section 7 gives our conclusions and some comments about 
the subtle limiting procedure required for computing back-reaction.

\section{The model and its solution}

The massive Schwinger model is defined by the Lagrangian density,
\begin{equation}
{\cal L}=\overline{\Psi} i\gamma^\mu (\partial_\mu+i e A_\mu)\Psi - m
\overline{\Psi} \Psi -{1\over 4} F_{\mu\nu} F^{\mu\nu} \; , \label{lagrangian}
\end{equation}
with $\mu, \nu=0,1$. $A_\mu$ is the gauge potential and $\Psi$ is the fermion 
Dirac doublet field. The electromagnetic coupling constant is $e < 0$. Our 
conventions for the flat spacetime metric, and the Gamma matrices are the ones 
used by Bjorken and Drell \cite{bd}, appropriately reduced to two spacetime 
dimensions. Thus, $\eta_{\mu\nu}=(+1, -1)$, $\{\gamma_\mu, \gamma_\nu\}=2\eta_{
\mu\nu}$, $\gamma^0 {\gamma^\mu}^\dagger \gamma^0 = \gamma^\mu$, ${\gamma_5
}^\dagger= \gamma_5$.

We shall adhere to the standard light-cone conventions \cite{Kogut} for 
defining coordinates, $x^{\pm}\equiv {1\over{\sqrt{2}}}(x^0\pm x^1)$. Identical
formulae relate the light-cone components of an arbitrary vector $V^\mu$ to the
Cartesian ones. In these coordinates the inner product of two vectors is $a^\mu
b_\mu=a^0 b^0-a^1b^1=a^+b^-+a^-b^+$. The non-vanishing components of the metric
are $g_{+-}=g_{-+}=1$, and consequently for any vector one has $a_+=a^-, a_-=
a^+$. With the standard definition $\partial_\pm\equiv \partial/\partial 
x^\pm$, one obtains $\partial_\mu V^\mu=\partial_+ V^+ + \partial_- V^-$ for 
any vector field $V^\mu$.

We define the light-cone spinor projection operators by ${\cal P}_{\pm}\equiv 
{1\over 2}(I\pm \gamma^0 \gamma^1)={1\over 2}(I\pm \gamma_5)={1\over 2}
\gamma^\mp \gamma^\pm$. In direct analogy with all other vectors $\gamma^\pm 
\equiv {1\over {\sqrt{2}}}(\gamma^0\pm \gamma^1)$. They satisfy $(\gamma^\pm)^2
=0$ and $\{\gamma^+,\gamma^-\}=2$. The Dirac spinor decomposes into $\Psi=
\psi_+ + \psi_-$, with $\psi_\pm \equiv {\cal P}_\pm \Psi$. Finally, note that 
in 1+1 dimensions ${\rm Tr}{\cal P}_{\pm} = 1$.

A convenient representation of the Dirac matrices is $\gamma^0=\sigma_1$, 
$\gamma^1=i\sigma_2$. In this representation $\gamma_5=\gamma^0 \gamma^1=-
\sigma_3$, while ${\cal P}_+={\rm diag}(0, 1)$ and ${\cal P}_-={\rm diag}(1, 
0)$. However, we shall work in a representation independent way.

We shall here go over the relevant bilinears without attention to either 
operator ordering or to regularization. These issues will be fully addressed
in Sections 5 and 6. The light-cone components $J^\pm\equiv {1 \over 
{\sqrt{2}}} (J^0\pm J^1)$ of the electromagnetic current $J^\mu=e\overline{
\Psi} \gamma^\mu \Psi$ are,
\be
J^\pm=\sqrt{2}e\psi_\pm^\dagger \psi_\pm \; . \label{J+-}
\ee
Similarly, the light-cone components $J_5^\pm\equiv {1\over{\sqrt{2}}}(J_5^0
\pm J_5^1)$ of the axial current $J_5^\mu=\overline{\Psi} \gamma^\mu\gamma_5
\Psi$ take the form,
\be
J_5^\pm=\pm\sqrt{2}\psi_\pm^\dagger \psi_\pm = \pm J^\pm /e \; .  \label{J5+-}
\ee
Finally, the bilinear $J_5\equiv \overline{\Psi} \gamma_5 \Psi$ in terms of 
$\psi_\pm$ becomes,
\be
J_5={1\over\sqrt{2}}(\psi_-^\dagger \gamma^+ \psi_+ - \psi_+^\dagger \gamma^- 
\psi_-) \; . \label{J5}
\ee

We fix the gauge by setting $A_+=0$ with the initial value of $A_-$ also zero,
\begin{equation}
A_-(x^+) = - \int_0^{x^+} du E(u) \; .
\end{equation}
The Dirac equation in an arbitrary $A_-(x^+)$ background field takes the form:
\begin{equation}
\left(\gamma^+ i \partial_+ + \gamma^- (i\partial_- - 
e A_-) - m \right) \Psi =0. \label{dirac}
\ee
It implies the conservation of the electromagnetic current
\begin{equation}
\partial_+ J^+ +\partial_- J^- =0 \; , \label{naive1}
\end{equation}
as well as the following relation between $J_5$ and the divergence of the
axial vector current,
\begin{equation}
e \partial_{\mu} J_5^{\mu} - 2 i e m J_5 = 2 \partial_+ J^+ - 2 i e m J_5 = 0 
\; . \label{naive2}
\end{equation}

\begin{figure}
\centerline{\epsfig{file=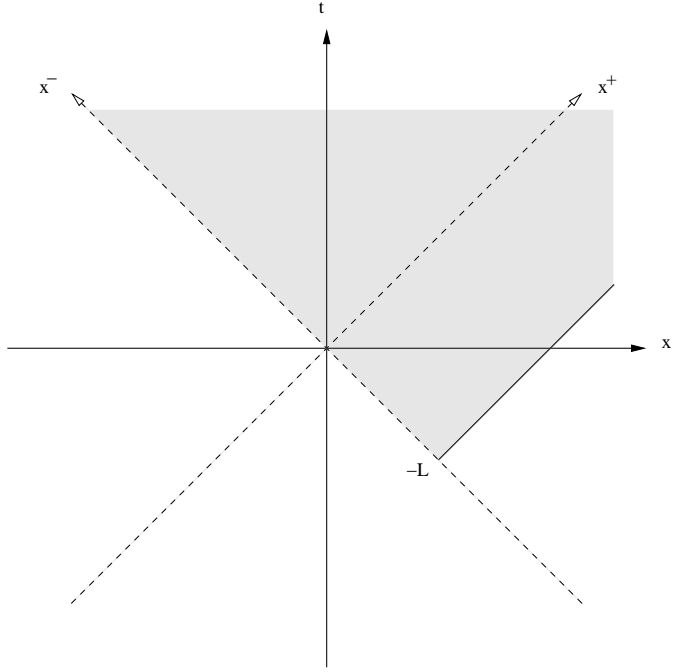,height=3.5in}}
\caption{The domain (shaded) of our solution for $\psi_{\pm}(x^+,x^-)$. We 
specify the initial values of $\psi_+(0,x^-)$ for all $x^- > -L$ and of
$\psi_-(x^+,-L)$ for all $x^+ > 0$.}
\end{figure}

Multiplying (\ref{dirac}) by $\gamma^+$ and $\gamma^-$ respectively, one may 
rewrite it in the equivalent form of the system of equations,
\begin{eqnarray}
i\partial_+ \psi_+ & = & {m\over 2}\gamma^-\psi_- \; , \label{eqn1} \\
(i\partial_- - eA_-)\psi_- & = & {m\over 2}\gamma^+\psi_+ \; . \label{eqn2}
\end{eqnarray}
Equations (\ref{eqn1}) and (\ref{eqn2}) can be integrated,
\begin{eqnarray}
\psi_+(x^+,x^-) & = & \psi_+(0,x^-)-{{im}\over 2}\int_0^{x^+} du \gamma^-
\psi_-(u,x^-) \label{int1} \; , \\
\psi_-(x^+,x^-) & = & e^{-ieA_-(x^+)(x^-+L)}\psi_-(x^+,-L) \nonumber \\
& & \qquad - {{im}\over 2} \int_{-L}^{x^-} dv e^{-ieA_-(x^+) (x^- - v)} 
\gamma^+\psi_+(x^+,v) \; , \label{int2}
\end{eqnarray}
to express $\psi_\pm(x^+,x^-)$ in terms of initial data $\psi_+(x^+=0,x^->-L)$ 
and $\psi_-(x^+>0,x^-=-L)$ on the semi-axes shown in Fig.~1.
Substituting (\ref{int2}) into (\ref{int1}) and iterating, one is led to the
following infinite series solution for $\psi_+$,
\begin{eqnarray}
\lefteqn{\psi_+(x^+, x^-)=\sum_{n=0}^{\infty}\Bigl(-{{m^2}\over 2}\Bigr)^n 
\int_0^{x^+}du_1 \int_{-L}^{x^-}dv_1 e^{-ieA_-(u_1) (x^- - v_1)} \int_0^{u_1}
du_2 } \nonumber \\
& & \int_{-L}^{v_1}dv_2 e^{-ieA_-(u_2) (v_1 - v_2)} \cdots \int_0^{u_{n-1}}du_n 
\int_{-L}^{v_{n-1}}dv_n e^{-ieA_-(u_n) (v_{n-1} - v_n)} \nonumber \\
& & \hspace{2cm} \times \Biggl\{\psi_+(0,v_n)- i{m\over 2}\gamma^- 
\int_0^{u_n}du e^{-ieA_-(u)(v_n+L)}\psi_-(u,-L)\Biggr\}. \quad\label{psi1}
\end{eqnarray}
 
One next introduces the identity,
\begin{equation}
\int_{-L}^{v_{n-1}} dv_n e^{-ieA_-(v_{n-1}-v_n)} = \int_{-L}^{\infty} dv_n 
\int_{-\infty}^\infty{{dk^+}\over{2\pi}} {{ie^{-i(k^+ + i/L)(v_{n-1} - v_n)}} 
\over {k^+ - eA_-(u_n)+i/L}} \; , \label{id}
\end{equation}
sets $v_n = v$ and interchanges the order of integration to get,
\begin{eqnarray}
\lefteqn{\psi_+(x^+, x^-)= \int_{-L}^{\infty} dv \int_{-\infty}^{\infty} {dk^+
\over 2\pi} e^{i (k^+ + i/L) v} \sum_{n=0}^{\infty} \Bigl(-{{m^2}\over 2}
\Bigr)^n \int_0^{x^+}du_1 \int_{-L}^{x^-}dv_1} \nonumber \\
& & \times e^{-ieA_-(u_1) (x^- - v_1)} \cdots \int_0^{u_{n-3}} du_{n-2}
\int_{-L}^{v_{n-3}} dv_{n-2} e^{-ie A_-(u_{n-2})(v_{n-3}-v_{n-2})} \nonumber \\
& & \qquad \times \int_0^{u_{n-2}} du_{n-1} \int_{-L}^{v_{n-2}}dv_{n-1} 
e^{-ie A_-(u_{n-1}) (v_{n-2} - v_{n-1})} e^{-i(k^+ + i/L)v_{n-1}} \nonumber \\
& & \qquad \qquad \times \int_0^{u_{n-1}} {i du_n \over{k^+ - eA_-(u_n)+i/L}}
\Biggl\{\psi_+(0,v) \nonumber \\
& & \hspace{4cm} - i{m\over 2}\gamma^- \int_0^{u_n} du 
e^{-ieA_-(u)(v+L)} \psi_-(u,-L)\Biggr\} \; . \qquad \label{psi2}
\end{eqnarray}
This is the key step at which we deviate from our previous solution \cite{ttw}.
In particular it will not be necessary, proceeding this way, to take $L$ to
infinity.

The integration over $v_{n-1}$ gives,
\begin{eqnarray}
\lefteqn{\int_{-L}^{v_{n-2}}dv_{n-1} e^{-ie A_-(u_{n-1}) (v_{n-2} - v_{n-1})}
e^{-i(k^+ +i/L)v_{n-1}} } \nonumber \\
& & \qquad = {i e^{-i(k^+ + i/L)v_{n-2}} - i e^{-i e A_-(u_{n-1}) (L + 
v_{n-2})} e^{i(k^+ +i/L)L} \over k^+ - eA_-(u_{n-1}) + i/L} \; . \label{vn1}
\end{eqnarray}
Note that the second term in (\ref{vn1}) gives zero inside the $k^+$ integral.
This is because $(v+L)$ is positive in the range over which $v$ is integrated,
so we can close the contour of the $k^+$ integral above. The result must be
zero because the integrand is analytic in the upper half plane.

Upon substitution of (\ref{vn1}) our solution for $\psi_+$ becomes,
\begin{eqnarray}
\lefteqn{\psi_+(x^+, x^-) = \int_{-L}^{\infty} dv \int_{-\infty}^{+\infty}
{{dk^+}\over{2\pi}} e^{i(k^+ + i/L) v} \sum_{n=0}^{\infty}\Bigl(-{{m^2}\over 2}
\Bigr)^n \int_0^{x^+}du_1 \dots}
\nonumber \\ 
& & \cdots \int_0^{u_{n-3}}du_{n-2} \int_{-L}^{v_{n-3}}dv_{n-2} 
e^{-ieA_-(u_{n-2}) (v_{n-3} - v_{n-2})} e^{-i(k^+ + i/L)v_{n-2}} \nonumber \\
& & \qquad \times \int_0^{u_{n-2}} {i du_{n-1} \over {k^+ - eA_-(u_{n-1})+i/L}} 
\int_0^{u_{n-1}} {i du_n \over {k^+ - eA_-(u_n)+i/L}} \nonumber \\
& & \hspace{2cm} \times \Biggl\{\psi_+(0,v) - i{m\over 2}\gamma^- 
\int_0^{u_n}du e^{-ieA_-(u)(v+L)}\psi_-(u,-L)\Biggr\} \; . \qquad \label{psi3}
\end{eqnarray}
Then perform the $v_{n-2}$ integration, just like (\ref{vn1}), and discard the 
lower limit as before. In this way all the $v_i$ integrations can be done to 
give,
\begin{eqnarray}
\lefteqn{\psi_+(x^+, x^-) = \int_{-L}^{\infty}dv \int_{-\infty}^{+\infty}
{dk^+\over 2 \pi} e^{i(k^+ + i/L) (v - x^-)}} \nonumber \\
& & \times \sum_{n=0}^{\infty} \int_0^{x^+} {-im^2/2 \; du_1 \over k^+ - 
eA_-(u_1) + i/L} \cdots \int_0^{u_{n-1}} {{-im^2/2 \; du_n} \over {k^+ - 
eA_-(u_n) + i/L}} \nonumber \\
& & \hspace{2cm} \times \Biggl\{ \psi_+(0,v) - {i\over 2} m \gamma^- 
\int_0^{u_n} du e^{-ieA_-(u)(v+L)} \psi_-(u,-L) \Biggr\} \; .\quad \label{psi4}
\end{eqnarray}

Use of the identities,
\begin{equation}
\int_0^{x^+} du_1 f(u_1) \int_0^{u_1} du_2 f(u_2) \cdots \int_0^{u_{
n-1}}du_n f(u_n) = {1 \over n!} \Biggl[\int_0^{x^+} dy f(y)\Biggr]^n \; ,
\end{equation}
\begin{eqnarray}
\lefteqn{\int_0^{x^+} du_1 f(u_1) \cdots \int_0^{u_{n-1}} du_n f(u_n) 
\int_0^{u_n} du g(u)} \nonumber \\
& & \hspace{4cm} = \int_0^{x^+} du g(u) {1\over{n!}} \Biggl[ \int_u^{x^+} dy
f(y)\Biggr]^n \; , 
\end{eqnarray}
leads to the final expression for $\psi_+$,
\begin{eqnarray}
\lefteqn{\psi_+(x^+,x^-) = \int_{-L}^{\infty}dv \int_{-\infty}^{+\infty} 
{{dk^+}\over{2\pi}} e^{i(k^++i/L) (v - x^-)} \Biggl\{{\cal E}[eA_-](0,x^+;k^+)
\psi_+(0,v)} \nonumber \\
& & \hspace{1cm} - {i\over 2} m \gamma^- \int_0^{x^+} du e^{-ieA_-(u)(v+L)} 
{\cal E}[eA_-](u,x^+;k^+) \psi_-(u,-L)\Biggr\} \; . \quad \label{psi+}
\end{eqnarray}
The functional ${\cal E}[eA_-]$ is,
\be
{\cal E}[eA_-](u,x^+;k^+) \equiv \exp\Biggl[- {i\over 2} m^2 \int_u^{x^+} 
{{du'} \over {k^+-eA_-(u')+i/L}}\Biggr] \; .  \label{w}
\ee
As for $\psi_-(x^+,x^-)$, it is obtained trivially from (\ref{eqn1}),
\begin{eqnarray}
\lefteqn{\psi_-(x^+,x^-) = {1\over m} \gamma^+ i \partial_+ \psi_+(x^+,x^-)
= e^{-ieA_-(x^+)(x^-+L)} \psi_-(x^+,-L)} \nonumber \\
& & + \int_{-L}^{\infty} dv \int_{-\infty}^{+\infty} {{dk^+}\over{2\pi}} {{m/2}
\; e^{i(k^++i/L) (v - x^-)} \over {k^+-eA_-(x^+)+i/L}}
\Biggl\{{\cal E}[eA_-](0,x^+;k^+) \gamma^+ \psi_+(0,v) \nonumber \\
& & \hspace{2cm} - i m \int_0^{x^+} du e^{-ieA_-(u)(v+L)} {\cal E}[eA_-](u,
x^+;k^+) \psi_-(u,-L)\Biggr\} \; . \qquad \label{psi-}
\end{eqnarray}

Equations (\ref{psi+}) and (\ref{psi-}) represent the general solution of the 
Dirac equation on the light-cone and in the presence of an arbitrary 
$x^+$ dependent background electric field $E(x^+)=-A'_-(x^+)$. The solutions
are expressed in terms of data given on the initial surface formed by the union
of the semi-axes $(x^+=0, x^->-L)$ and $(x^+>0, x^-=-L)$, shown in Fig~1. It is
straightforward to verify that our solutions for $\psi_{\pm}$ obey the Dirac 
equations (\ref{eqn1}-\ref{eqn2}). It is also obvious from (\ref{psi+}) that
our solution for $\psi_+$ agrees with its initial value at $x^+ = 0$. To see
that our solution for $\psi_-$ agrees with its initial value at $x_- = -L$ note
the second term in (\ref{psi-}) gives zero (at $x^- = -L$ only!) for the same 
reason that the lower limit of (\ref{vn1}) makes no contribution. When $x^- = 
-L$ the term $(v - x^-)$ is positive and we can close the $k^+$ contour above. 
Since the integrand is analytic in the upper half plane the result is zero.

\section{Quantization} 

Our solutions (\ref{psi+}) and (\ref{psi-}) depend upon $\psi_+(0,v)$ and
$\psi_-(u,-L)$. We use canonical quantization to define the algebra of these 
initial value operators. The Lagrangian for this system is,
\begin{eqnarray}
\lefteqn{{\cal L}={\overline{\Psi}}(i\gamma^\mu \partial_\mu -e\gamma^\mu 
A_\mu-m)\Psi \; ,} \\
& & = \sqrt{2}\psi_+^\dagger\Bigl(i\partial_+\psi_+ -{m\over 2}\gamma^- 
\psi_-\Bigr) + \sqrt{2}\psi_-^\dagger\Bigl((i\partial_- - eA_-)\psi_- - {m\over
2}\gamma^+ \psi_+\Bigr) \; . \quad 
\end{eqnarray}
The momentum conjugate to $\psi_+(0,v)$ is $i\sqrt{2}\psi_+^\dagger(0,v)$, the 
partial derivative of the Lagrangian density with respect to the normal 
derivative $\partial_+\psi_+$ evaluated on the $x^+=0$ branch of the initial 
Cauchy surface. Correspondingly, the momentum conjugate to $\psi_-(u,-L)$
is the partial derivative of ${\cal L}$ with respect to the normal derivative
$\partial_-\psi_-$ on the $x^-=-L$ branch, and is equal to $i\sqrt{2}\psi_-^{
\dagger}(u,-L)$. Because the two branches are spacelike separated the canonical 
coordinates and momenta defined on them must be independent of one another. 
Hence the only non-zero anti-commutators are,
\begin{eqnarray}
\Bigl\{\psi_+(0,v),\psi_+^\dagger(0,v')\Bigr\} & = & {1\over\sqrt{2}}{\cal P}_+
\delta(v-v') \; , \label{facr1} \\
\Bigl\{\psi_-(u,-L),\psi_-^\dagger(u',-L)\Bigr\} & = & {1\over\sqrt{2}}
{\cal P}_- \delta(u-u') \; . \label{facr2}
\end{eqnarray}

The algebra of the initial value operators, plus the way in which our solutions
(\ref{psi+},\ref{psi-}) depend upon the initial value operators, determines
how the various operators anti-commute at any spacetime point. It is 
straightforward to check that the expected equal $x^+$ and equal $x^-$ 
relations do in fact result,
\begin{eqnarray}
\Bigl\{\psi_+(x^+,x^-),\psi_+^\dagger(x^+,y^-)\Bigr\} & = & {1\over\sqrt{2}}
{\cal P}_+ \delta(x^- -y^-) \; , \\
\Bigl\{\psi_-(x^+,x^-),\psi_-^\dagger(y^+,x^-)\Bigr\} & = & {1\over\sqrt{2}}
{\cal P}_- \delta(x^+ -y^+) 
\end{eqnarray}
Note that these relations would {\it not} have followed if we had adopted the
usual light-cone procedure of ignoring dependence upon $\psi_-(x^+,-L)$. 
Properly resolving the ambiguity at $k^+ = 0$ is therefore necessary to 
preserve unitarity.

\vfill\eject

It is convenient to define the ``Fourier transform'' of $\psi_+$ as follows,
\begin{eqnarray}
\lefteqn{\widetilde{\psi}_+(x^+,k^+) \equiv \int_{-L}^{\infty} dv e^{i(k^++i/L)
v} \Biggl\{{\cal E}[eA_-](0,x^+;k^+) \psi_+(0,v)} \nonumber \\
& & \hspace{1cm} - {i\over 2} m \gamma^- \int_0^{x^+} du e^{-ieA_-(u)(v+L)} 
{\cal E}[eA_-](u,x^+;k^+) \psi_-(u,-L)\Biggr\} \; . \quad \label{psitilde}
\end{eqnarray}
In the large $L$ limit this quantity becomes a pure creation or annihilation
operator. To see why, first note that $x^+$ is the light-cone time parameter. 
Now compute the $x^+$ derivative of $\widetilde{\psi}_+$,
\begin{eqnarray}
\lefteqn{-i\partial_+ \widetilde{\psi}_+(x^+,k^+) = {-{m^2/2} \; 
\widetilde{\psi}_+(x^+,k^+) \over {k^+-eA_-(x^+)+i/L}}} \nonumber \\
& & \hspace{3cm} - {{i m/2} \; e^{-i(k^++i/L)L} \over {k^+ - eA_-(x^+) + i/L}} 
\gamma^- \psi_-(x^+,-L) \; . \label{d+psi}
\end{eqnarray}
In the large L limit, the second term contributes only for $k^+ = eA_-(x^+)$
because,
\begin{equation}
\lim_{L \rightarrow \infty} {{e^{-i(k^+ - e A_-(x^+) + i/L)L}} \over 
{k^+ - eA_-(x^+) + i/L}} = -2 \pi i \delta\Bigl(k^+ - e A_-(x^+)\Bigr) \; .
\end{equation}
Therefore, away from the singular point at $k^+ = e A_-(x^+)$ and in the large
$L$ limit, $\widetilde{\psi}_+(x^+,k^+)$ is an eigenoperator of the light-cone
Hamiltonian. For $k^+ < e A_-(x^+)$ its eigenvalue is positive so it must
create a positron with some amplitude. For $k^+ > e A_-(x^+)$ its eigenvalue
is negative so it must annihilate an electron with some amplitude.

To find the amplitude we compute the anti-commutator between the operator and
its conjugate,
\begin{eqnarray}
\lefteqn{\{\widetilde{\psi}_+(x^+,k^+),\widetilde{\psi}_+^\dagger(x^+,q^+)\} =
\int_{-L}^{\infty}dv e^{i(k^++i/L)v} \int_{-L}^{\infty}dw e^{-i(q^+-i/L)w} }
\nonumber \\
& & \quad \times {{\cal P}_+\over\sqrt{2}} \Biggl\{{\cal E}(0,x^+;k^+){\cal 
E}^*(0,x^+;q^+) \delta(v-w) +{m^2 \over 2} \int_0^{x^+}due^{-ieA_-(u)(v+L)} 
\nonumber \\
& & \qquad \qquad \times {\cal E}(u,x^+;k^+) \int_0^{x^+} dy e^{ieA_-(y)(w+L)} 
{\cal E}^*(y,x^+;q^+) \delta(u-y)\Biggr\} \; , \\
& & = {1\over\sqrt{2}} {\cal P}_+ {{ie^{-i(k^+-q^++2i/L)L}}\over{k^+-q^++2i/L}}
\Biggl\{{\cal E}(0,x^+;k^+) {\cal E}^*(0,x^+;q^+) \nonumber \\
& & \hspace{3cm} + \int_0^{x^+} du {\partial \over {\partial u}} 
\Bigl({\cal E}(u,x^+;k^+){\cal E}^*(u,x^+;q^+)\Bigr) \Biggr\} \; , \\
& & = {1\over\sqrt{2}} {\cal P}_+ {{ie^{-i(k^+-q^++2i/L)L}} \over {k^+ - q^+ +
2 i/L}} \; , \\
& & \longrightarrow {1\over\sqrt{2}} {\cal P}_+ 2\pi \delta(k^+-q^+) \; .
\label{anti}
\end{eqnarray}
We conclude that, in the limit $L\to \infty$, $2^{1/4}\tilde\psi_+(x^+,k^+)$ 
creates positrons with unit amplitude for $k^+<eA_-(x^+)$ and destroys 
electrons with unit amplitude for $k^+>eA_-(x^+)$.

It remains to specify the vacuum. Of course a system for which pair production
occurs is not stable and so does not possess a true ground state. It is 
nevertheless reasonable to work with the state $\vert \Omega \rangle$ which is
empty on the initial value surface. That is, $\vert \Omega \rangle$ is the
usual vacuum for Dirac theory with $A_- = 0$. Since we are in the Heisenberg 
picture $\vert \Omega \rangle$ is the state of the system for all times but
a nonzero background shows up in how the field operators depend upon their
initial values. Thus our method for computing the expectation value of an 
operator at $(x^+,x^-)$ is to use (\ref{psi+},\ref{psi-}) to reduce the 
problem to expectation values of the initial value operators. These are then
computed in the well-known $A_- = 0$ theory. For example, the expectation
value of any bilinear can be read off from the following \cite{bd}:
\begin{eqnarray}
\lefteqn{\Bigl\langle \Omega \Bigl\vert \psi_\alpha(x^+,x^-) 
\psi_\beta^\dagger(y^+,y^-) \Bigr\vert \Omega \Bigr\rangle_{A_- = 0}} 
\nonumber \\
& & \hspace{2cm} = \int_{-\infty}^\infty{{dp}\over{2\pi}} (\sla{p} \gamma^0 
+ m\gamma^0)_{ \alpha\beta} {1 \over 2 \omega} e^{-ip^-(x^+ -y^+) - i p^+
(x^- -y^-)} \; , \label{vev1} \\
\lefteqn{\Bigl\langle \Omega \Bigl\vert \psi_\beta^\dagger(y^+,y^-) 
\psi_\alpha(x^+,x^-) \Bigr\vert \Omega \Bigr\rangle_{A_- = 0}} \nonumber \\
& & \hspace{2cm} = \int_{-\infty}^\infty{{dp}\over{2\pi}} (\sla{p} \gamma^0 
- m\gamma^0)_{\alpha\beta} {1 \over 2 \omega} e^{ip^-(x^+ -y^+) + i p^+
(x^- -y^-)} \; . \label{vev2}
\end{eqnarray}
In these integrals the light-cone momenta are given by $p^\pm = {1 \over 
\sqrt{2}} (\omega \pm p)$, with $\omega=\sqrt{m^2+p^2}$ and $2p^+ p^- = m^2$. 
It is often convenient to change variables from $p$ to $p^+$ or $p^-$,
\be
\int_{-\infty}^\infty dp = \int_0^\infty dp^+ {\omega\over {p^+}} 
= \int_0^\infty dp^- {\omega\over {p^-}} \; .
\ee

We can drop the subscript ``$A_- = 0$'' when the coordinates $(x^+,x^-)$ and 
$(y^+,y^-)$ are specialized to the initial value surface because the state 
$\vert \Omega \rangle$ is defined to agree with the $A_- = 0$ vacuum on this 
surface. For example, taking the spinor trace of the various $\pm$ components
of (\ref{vev2}) gives,
\begin{eqnarray}
\Bigl\langle \O \Bigl\vert \psi_+^\dagger (0,z) \psi_+(0,v) \Bigr\vert \O 
\Bigr\rangle & = & {1\over{\sqrt{2}}} \int_{-\infty}^0 {{dp^+} \over {2\pi}} 
e^{-ip^+(v-z)} \; , \\
\Bigl\langle \O \Bigl\vert \psi_+^\dagger (0,z)\gamma^-\psi_-(y,-L) \Bigr\vert 
\O \Bigr\rangle & = & {1\over{\sqrt{2}}} \int_{-\infty}^0 {{dp^+}\over{2\pi}}
{m\over {p^+}}e^{-ip^-u+ip^+(z+L)} \; , \\
\Bigl\langle \O \Bigl\vert \psi_-^\dagger (y,-L)\gamma^+\psi_+(0,v) \Bigr\vert
\O \Bigr\rangle & = & {1\over{\sqrt{2}}}\int_{-\infty}^0{{dp^+}\over{2\pi}}
{m\over {p^+}}e^{ip^-y-ip^+(v+L)} \; , \\
\Bigl\langle \O \Bigl\vert \psi_-^\dagger (y,-L)\psi_-(u,-L) \Bigr\vert \O
\Bigr\rangle & = &
{1\over{\sqrt{2}}}\int_{-\infty}^0{{dp^-}\over{2\pi}}e^{-ip^-(u-y)} \; , \\
& = & {1\over{\sqrt{2}}}\Biggl[{1\over 2}\delta(u-y)+
{i\over{2\pi}}{\cal P}\Bigl({1\over{u-y}}\Bigr)\Biggr] \; . \label{formulas}
\end{eqnarray}
Note that in taking $L$ to infinity we can neglect the cross correlators
between $\psi_+(0,x^-)$ and $\psi_-(x^+,-L)$.

\section{Pair production on the light-cone}

It was shown above that the $L \rightarrow \infty$ limit of the operator 
$2^{1/4} \widetilde{\psi}_+(x^+,k^+)$ has unit amplitude for creating positrons
when $k^+<eA_-(x^+)$ and for destroying electrons when $k^+>eA_-(x^+)$. Now
recall that the vector potential is minus the integral of the electric field,
\begin{equation}
A_-(x^+) = - \int_0^{x^+} du E(u) \; .
\end{equation}
Since the electron charge $e$ is negative we see that the function $e A_-(x^+)$
increases monotonically from zero at $x^+ = 0$ for as long as the electric 
field remains positive. Therefore modes with positive $k^+$ start out as 
electron annihilation operators and then become positron creators after the 
critical time $x^+ = X(k^+)$ at which $e A_-(x^+) = k^+$. This is the 
phenomenon of pair creation. 

Two important qualitative facts deserve mention although they were both 
explained in our previous paper \cite{ttw}. The first is that, on the 
light-cone, pair creation is an instantaneous and singular event. The second
is that the newly created $e^-$ instantly leaves the manifold, so we see only
the $e^+$. Both facts are explained by noting that the light-cone problem of 
evolving a state from $x^+ = 0$ corresponds to the infinite boost limit of the 
conventional problem of evolving a state from $x^{0\prime} = 0$ in the primed
frame \cite{Kogut}. In the latter problem pair creation dribbles out, a little 
at a time, for modes of all different momenta. However, it is straightforward 
to show that any particle created with finite primed momentum will have $p^+ = 
0$ after the infinite boost. In a background gauge field the physical momentum 
is the minimally coupled one, $p^+ = k^+ - eA_-(x^+)$. So $p^+ = k^+ - e 
A_-(x^+) = 0$ defines the instant of pair creation in the light-cone problem. 
Electrons immediately leave the light-cone manifold because, in the primed 
frame they accelerate {\it opposite} to the direction of the electric field, 
which is also the direction of the boost. Electrons therefore emerge, in the 
light-cone problem, moving at the speed of light along the negative $x^-$ axis,
which takes them off the light-cone manifold immediately. Positrons emerge 
moving at the speed of light parallel to the $x^+$ axis, so they remain on the 
manifold. The process is depicted in Fig.~2.

\begin{figure}
\centerline{\epsfig{file=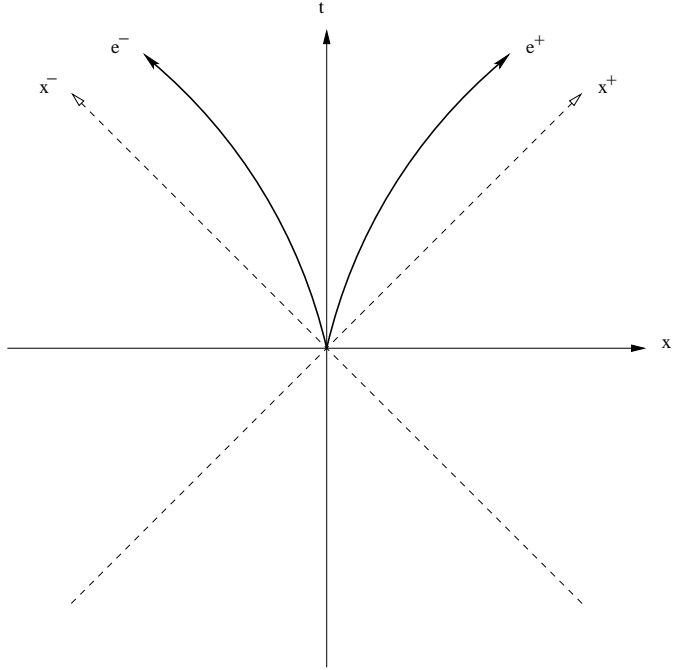,height=3.5in}}
\caption{The evolution of an $e^+ e^-$ pair. Note that the electron does not
appear beyond a certain value of $x^+$.}
\end{figure}

It remains to compute the probability for pair creation Prob($e^+$) at time 
$x^+ > X(k^+)$. From the previous discussion this should be given by the 
relation,
\be
\lim_{L\to\infty} \sqrt{2} \Bigl\langle \Omega \Bigl\vert \widetilde{
\psi}_+^\dagger(x^+,q^+) \, \widetilde{\psi}_+(x^+,k^+) \Bigr\vert \Omega 
\Bigr\rangle = \Bigl[1-{\rm Prob}(e^+) \Bigr] 2\pi \delta(k^+-q^+) \; .
\ee
To take the distributional limit rigorously we first smear with test functions
$f^*(k^+)$ and $f(q^+)$,
\begin{eqnarray}
\lefteqn{\int_{-\infty}^{+\infty} {{dk^+} \over {2\pi}} f^*(k^+) \int_{-\infty
}^{+\infty} {{dq^+} \over {2\pi}} f(q^+) \sqrt{2} \Bigl\langle \Omega 
\Bigl\vert \widetilde{\psi}_+^\dagger(x^+,q^+) \, \widetilde{\psi}_+(x^+,k^+) 
\Bigr\vert \Omega \Bigr\rangle =} \nonumber \\
& & \sqrt{2} \int_{-\infty}^{+\infty} {{dk^+} \over {2\pi}} f^*(k^+)
\int_{-\infty}^{+\infty} {{dq^+} \over {2\pi}} f(q^+) \int_{-L}^\infty dv 
e^{i(k^++i/L)v} \int_{-L}^\infty dz e^{-i(q^+-i/L)z} \nonumber \\
& & \times \Biggl\{{\cal E}(0,x^+;k^+) {\cal E}^*(0,x^+;q^+) \Bigl\langle \O
\Bigl\vert \psi_+^\dagger(0,z) \psi_+(0,v) \Bigr\vert \O \Bigr\rangle 
\nonumber \\
& & -{i m \over 2} \int_0^{x^+} du e^{-ieA_-(u)(v+L)} {\cal E}(u,x^+;k^+)
{\cal E}^* \Bigl\langle \O \Bigl\vert \psi_+^\dagger(0,z) \gamma^-
\psi_-(u,-L) \Bigr\vert \O \Bigr\rangle \nonumber \\
& & +{i m \over 2} \int_0^{x^+} dy e^{ieA_-(y)(z+L)} {\cal E} {\cal E}^*(y,
x^+;q^+) \Bigl\langle \O \Bigl\vert \psi_-^\dagger(y,-L) \gamma^+ \psi_+(0,v) 
\Bigr\vert \O \Bigr\rangle \nonumber \\
& & +{{m^2} \over 2} \int_0^{x^+} du e^{-ieA_-(u)(v+L)} {\cal E}(u,x^+;k^+)
\int_0^{x^+} dy e^{ieA_-(y)(z+L)} {\cal E}^*(y,x^+;q^+) \nonumber \\
&& \hspace{6cm} \times \Bigl\langle \O \Bigl\vert \psi_-^\dagger(y,-L) 
\psi_-(u,-L) \Bigr\vert \O \Bigr\rangle \Biggr\} \; . \label{prob1}
\end{eqnarray}

As noted in the previous section, the $\pm$ cross correlators vanish in the 
large $L$ limit. Therefore only the first ($++$) and the fourth ($--$) terms 
in (\ref{prob1}) make a non-vanishing contribution in the $L\to\infty$ limit.
The $++$ term gives,
\begin{eqnarray}
\lefteqn{\hspace{1cm} \int_{-\infty}^{\infty} {{dk^+} \over {2\pi}} f^*(k^+)
\int_{-\infty}^{\infty} {{dq^+} \over {2\pi}} f(q^+) \int_{-L}^\infty dv 
e^{i(k^++i/L)v} \int_{-L}^\infty dz e^{-i(q^+-i/L)z}} \nonumber \\ 
& & \hspace{3cm} \times \int_{-\infty}^0 {{dp^+} \over {2\pi}} e^{-ip^+(v-z)}
{\cal E}(0,x^+;k^+) {\cal E}^*(0,x^+;q^+) \; , \qquad \\
& & = \int_{-\infty}^0 {{dp^+} \over {2\pi}} \int_{-\infty}^{+\infty} {{dk^+}
\over {2\pi}} {{e^{-i[k^+-p^++i/L]L}} \over {k^+-p^++i/L}} {\cal E}(0,x^+;k^+)
f^*(k^+) \nonumber \\
& & \hspace{3cm} \times \int_{-\infty}^{+\infty} {{dq^+} \over{2\pi}} {{e^{i
(q^+ - p^+ -i/L)L}} \over {q^+-p^+-i/L}} {\cal E}^*(0,x^+;q^+)f(q^+) \; , 
\qquad \\
& & = \int_{-\infty}^0 {{dp^+} \over {2\pi}} \int_{-\infty}^{+\infty} {{da}
\over {2\pi}} {{e^{-i(a+i)}} \over {a+i}} {\cal E}(0,x^+;p^++a/L) f^*(p^++a/L)
\nonumber \\
& & \hspace{2cm} \times \int_{-\infty}^{+\infty} {{db} \over {2\pi}} {{e^{i
(b-i)}} \over {b-i}} {\cal E}^*(0,x^+;p^++b/L) f(p^+ + b/L) \; , \\
& & \longrightarrow \int_{-\infty}^0 {{dp^+} \over {2\pi}} \Bigl\Vert f(p^+)
\Bigr\Vert^2 \int_{-\infty}^{+\infty} {{da} \over {2\pi}} {{e^{-i(a+i)}} \over 
{a+i}} \int_{-\infty}^{+\infty} {{db} \over {2\pi}} {{e^{i (b-i)}} \over {b-i}}
\; , \\
& & = \int_{-\infty}^0{{dp^+} \over{2\pi}} \Bigr\Vert f(p^+) \Bigr\Vert^2 \; .
\label{term1}
\end{eqnarray}

Before reducing the $--$ term of (\ref{prob1}) it is useful to recall the
integral \cite{ttw},
\begin{equation}
\int_{-\infty}^{\infty} {da \over 2\pi} {e^{-i (a+i)} \over a + i} \exp\Bigl[
-i x \Bigl(\ln(a+i) - i \pi\Bigr) \Bigr] = {-i e^{-\pi x/2} \over \Gamma(1 +
i x)} \; , \label{aint}
\end{equation}
and the identity due to Lobachevskiy \cite{Grad},
\begin{equation}
{1 \over \Gamma(1 + i x) \Gamma(1 - i x)} = {\sinh(\pi x) \over \pi x} \; ,
\label{Loba}
\end{equation}
both valid for real, positive $x$. We must also introduce the function,
\begin{equation}
\lambda(p^+) \equiv {m^2 \over 2 \vert e \vert E(X(p^+))} \; . \label{lambda}
\end{equation}
The large $L$ limit of the $--$ term proceeds in a similar fashion to that of 
the $++$ term. The steps are,
\begin{eqnarray}
\lefteqn{ {{m^2} \over 2} \int {{dk^+} \over {2\pi}} f^*(k^+) \int {{dq^+}
\over {2\pi}} f(q^+) \int_{-L}^\infty dv e^{i(k^++i/L)v} \int_{-L}^\infty dz 
e^{-i(q^+-i/L)z}} \nonumber \\
& & \times \int_0^{x^+} du e^{-ieA_-(u)(v+L)} {\cal E}(u,x^+;k^+) \int_0^{x^+} 
dy e^{ieA_-(y)(z+L)} {\cal E}^*(y,x^+;q^+) \nonumber \\
& & \times \Biggl[{1\over 2} \delta(u-y) + {i\over{2\pi}} {\cal P}\Bigl({1 
\over {u-y}} \Bigr)\Biggr] = {{m^2}\over 2} \int_0^{x^+} du \int_0^{x^+} dy 
{i \over {2\pi}} {\cal P}\Bigl({1 \over u - y}\Bigr) \nonumber \\
& & \hspace{1cm} \times \int_{-\infty}^{\infty} {{dk^+} \over {2\pi}} 
{e^{-i(k^+ + i/L)L} \over {k^+-eA_-(u)+i/L}} {\cal E}(u,x^+;k^+) f^*(k^+) 
\nonumber \\
& & \hspace{1.5cm} \times \int_{-\infty}^{\infty} {dq^+ \over 2 \pi} {e^{i (q^+
- i/L)L} \over q^+ - eA_-(y) - i/L} {\cal E}^*(y,x^+;q^+) f(q^+) \nonumber \\
& & + {{m^2} \over 4} \int_0^{x^+} du \int_{-\infty}^{\infty} {{dk^+} \over {2
\pi}} {e^{-i(k^++i/L)L} \over {k^+-eA_-(u)+i/L}} {\cal E}(u,x^+;k^+) f^*(k^+) 
\nonumber \\
& & \hspace{2 cm} \times \int_{-\infty}^{\infty} {{dq^+} \over {2\pi}} {e^{i
(q^+-i/L)L} \over {q^+-eA_-(u)-i/L}} {\cal E}^*(u,x^+;q^+)f(q^+) \; , \\
& & = {{m^2} \over 2} \int_0^{x^+} du \int_{-(eA_-(x^+) - eA_-(u))L}^{eA_-(u)L}
{{dc} \over {2\pi}} {i\over L} {\cal P}\Biggl({ X'(eA_-(u) - c/L) \over 
u - X(eA_-(u) - c/L)}\Biggr) \nonumber \\
& & \times \int_{-\infty}^\infty {{da} \over {2\pi}} {e^{-i(a+i)} 
\over {a+i}} {\cal E}\Bigl(u,x^+;eA_-(u) + \frac{a}{L}\Bigr) f^*\Bigl(eA_-(u) + 
\frac{a}{L}\Bigr) \int_{-\infty}^\infty {{db} \over {2\pi}} {e^{i(b-c-i)}
\over {b-i}} \nonumber \\
& & \hspace{.5cm} \times {\cal E}^*\Biggl(X\Bigl(eA_-(u) - \frac{c}{L}\Bigr),
x^+;eA_-(u) + \frac{b - c}{L}\Biggr) f\Bigl(eA_-(u) + \frac{b-c}{L}\Bigr) 
\nonumber \\
& & + {{m^2} \over 4} \int_0^{x^+} du \int_{-\infty}^\infty {{da} \over {2\pi}}
{e^{-i(a+i)} \over {a+i}} {\cal E}\Bigl(u,x^+;eA_-(u) + \frac{a}{L}\Bigr) 
f^*\Bigl(eA_-(u) + \frac{a}{L}\Bigr) \nonumber \\
& & \hspace{1cm} \times \int_{-\infty}^\infty {{db} \over {2\pi}} {e^{i(b-i)}
\over {b-i}} {\cal E}^*\Bigl(u,x^+;eA_-(u) + \frac{b}{L}\Bigr) f\Bigl(eA_-(u) +
\frac{b}{L}\Bigr) \; , \\
& & \longrightarrow {{m^2} \over 2} \int_0^{x^+} du \Bigl\Vert f\Bigl(e A_-(u)
\Bigr) \Bigr\Vert^2 \int_{-\infty}^\infty {{da} \over {2\pi}} {e^{-i(a+i)}
\over {a+i}} e^{-i \lambda\Bigl(eA_-(u)\Bigr) \Bigl(\ln(a+i)-i\pi\Bigr)}
\nonumber \\
& & \hspace{.5cm} \times \int_{-\infty}^\infty {{db} \over {2\pi}} {e^{i(b-i)} 
\over {b-i}} e^{i \lambda\Bigl(eA_-(u)\Bigr) \Bigl(\ln(b-i)+i\pi\Bigr) }
\Biggl\{\frac12 + \int_{-\infty}^{\infty} {{dc} \over {2\pi}} {\cal P}\Bigl(
{i e^{-ic} \over c}\Bigr) \Biggr\} \; , \\
& & = {{m^2} \over 2} \int_0^{x^+} du \Bigl\Vert f\Bigl(eA_-(u)\Bigr) 
\Bigr\Vert^2 {{-ie^{-\pi \lambda(eA_-(u))/2}} \over {\Gamma(1 + i \lambda(eA_-(
u)))}} \; {{ie^{-\pi \lambda(eA_-(u))/2}} \over {\Gamma(1 - i \lambda(eA_-(u)))
}} \; , \\
& & = \int_0^{x^+} du \Bigl\Vert f\Bigl(eA_-(u)\Bigr) \Bigr\Vert^2 \Bigl[1 -
e^{-2\pi\lambda(eA_-(u))}\Bigr] \frac1{2\pi} eA'_-\Bigl(X(eA_-(u))\Bigr) \; ,\\
& & = \int_0^{eA_-(x^+)} {{dp^+} \over {2\pi}} \Bigl\Vert f(p^+) \Bigr\Vert^2
\Bigl[1-e^{-2\pi\lambda(p^+)}\Bigr] \; . \label{term4}
\end{eqnarray}

Combining (\ref{term1}) and (\ref{term4}) above we obtain the following
relation for the probability of $\vert \Omega \rangle$ containing a positron
of momentum $p^+$ at time $x^+$,
\begin{eqnarray}
\lefteqn{\int_{\infty}^{\infty} {dp^+ \over 2\pi} \Bigl\Vert f(p^+) 
\Bigr\Vert^2 \Bigl[1 - {\rm Prob}(e^+) \Bigr] = \int_{\infty}^0 {dp^+ \over 2 
\pi} \Bigl\Vert f(p^+) \Bigr\Vert^2} \nonumber \\
& & \hspace {4cm} + \int_0^{eA_-(x^+)} {{dp^+} \over {2\pi}} \Bigl\Vert 
f(p^+) \Bigr\Vert^2 \Bigl[1-e^{-2\pi\lambda(p^+)}\Bigr] \; . 
\end{eqnarray}
This means that the state remains empty for all $p^+ < 0$ but, for $0 < p^+ <
e A_-(x^+)$, it contains a positron with probability,
\be
{\rm Prob}(e^+) = e^{-2\pi\lambda(p^+)} \; . \label{probability}
\ee
It should be noted that this result was derived under the assumption that
$e A_-(x^+)$ increases monotonically. 

\section{The vector current}

We regulate the various fermion bilinears by gauge invariant point splitting.
It suffices to split the two currents $J^\pm$ along the directions $x^\mp$,
respectively,
\begin{eqnarray}
J^+(x^+;x^-,y^-) & \equiv & {e\over{\sqrt{2}}} e^{ie A_-(x^+) (x^- - y^-)}
\Biggl\{\psi_+^{\dagger}(x^+,y^-) \psi_+(x^+,x^-) \nonumber \\
& & \hspace{2.5cm} - {\rm Tr}\Bigl[\psi_+(x^+,x^-) \psi_+^{\dagger}(x^+,y^-)
\Bigr]\Biggr\} \; , \label{J+} \\
J^-(x^+,y^+;x^-) & \equiv & {e\over{\sqrt{2}}} \Bigl\{\psi_-^{\dagger}(y^+,x^-)
\psi_-(x^+,x^-) \nonumber \\
& & \hspace{2.5cm} - {\rm Tr}\Bigl[\psi_-(x^+,x^-) \psi_-^{\dagger}(y^+,x^-)
\Bigr] \Biggr\} \; .  \label{J-}
\end{eqnarray}
Although these quantities are well regulated and gauge invariant, they are not
yet Hermitian. We enforce Hermiticity by taking the symmetric average,
\begin{eqnarray}
J^+_S(x^+;x^-,y^-) & \equiv & {1\over 2} \Bigl\{J^+(x^+;x^-,y^-) +
J^+(x^+;y^-,x^-)\Bigr\} \; , \label{J+s} \\
J^-_S(x^+;x^-,y^-) & \equiv & {1\over 2} \Bigl\{J^-(x^+,y^+;x^-) +
J^-(y^+,x^+;x^-)\Bigr\} \; . \label{J-s}
\end{eqnarray}
This could have been done in a single step but it is somewhat more efficient, 
calculationally, to first compute the expectation values of $J^{\pm}$ in the 
large $L$ limit and then take the Hermitian average as we also remove the 
splitting.

Let us begin with $J^+$. As explained at the end of Section 3, the first step
consists of using our general solution (\ref{psi+}) to express the expectation
value of $J^+(x^+;x^- + \Delta,x^-)$ as a sum of correlation functions of the
initial value operators,
\vfill\eject

\begin{eqnarray}
\lefteqn{\Bigl\langle \Omega \Bigl\vert J^+(x^+;x^- + \Delta, x^-) \Bigr\vert
\Omega \Bigr\rangle = {e\over\sqrt{2}} e^{ieA_-(x^+)\Delta} \int_{-\infty
}^\infty {{dk^+} \over {2\pi}} e^{-i(k^++i/L) (x^- + \Delta)}} \nonumber \\
& & \times \int_{-L}^\infty dv e^{i(k^++i/L)v} \int_{-\infty}^\infty {{dq^+}
\over {2\pi}} e^{i(q^+-i/L)x^-} \int_{-L}^\infty dw e^{-i(q^+-i/L)w} 
\Biggl\{{\cal E}(0,x^+;k^+) \nonumber \\
& & \quad \times {\cal E}^*(0,x^+;q^+) \Bigl\langle \Omega \Bigl\vert 
\psi_+^\dagger(0,w) \psi_+(0,v) - {\rm Tr}\Bigl[\psi_+(0,v) \psi_+^\dagger(0,w)
\Bigr] \Bigr\vert \Omega \Bigr\rangle \nonumber \\
& & + {{im} \over 2} {\cal E}(0,x^+;k^+) \int_0^{x^+} dy e^{ieA_-(y)(w+L)}
{\cal E}^*(y,x^+;q^+) \nonumber \\
& & \hspace{1.5cm} \times \Bigl\langle \Omega \Bigl\vert \psi_-^\dagger(y,-L) 
\gamma^+ \psi_+(0,v) - {\rm Tr}\Bigl[\psi_+(0,v) \psi_-^\dagger(y,-L) \gamma^+ 
\Bigr] \Bigr\vert \Omega \Bigr\rangle \nonumber \\
& & -{{im} \over 2} \int_0^{x^+} du e^{-ieA_-(u)(v+L)} {\cal E}(u,x^+;k^+)
{\cal E}^*(0,x^+;q^+) \nonumber \\
& & \hspace{1.5cm} \times \Bigl\langle \Omega \Bigl\vert \psi_+^\dagger(0,w) 
\gamma^- \psi_-(u,-L) - {\rm Tr}\Bigl[\gamma^- \psi_-(u,-L) \psi_+^\dagger(0,w)
\Bigr] \Bigr\vert \Omega \Bigr\rangle \nonumber \\
& & + {{m^2} \over 2} \int_0^{x^+} du e^{-ieA_-(u)(v+L)} {\cal E}(u,x^+;k^+)
\int_0^{x^+} dy e^{ieA_-(y)(w+L)} {\cal E}^*(y,x^+;q^+) \nonumber \\
& & \hspace{1cm} \times \Bigl\langle \Omega \Bigl\vert \psi_-^\dagger(y,-L)
\psi_-(u,-L) - {\rm Tr}\Bigl[\psi_-(u,-L) \psi_-^\dagger(y,-L) \Bigr] 
\Bigr\vert \Omega \Bigr\rangle \Biggr\} . \quad \label{<J+>}
\end{eqnarray}

The various correlation functions which appear in (\ref{<J+>}) can be read off
from relations (\ref{vev1}) and (\ref{vev2}). First note that the spinor
identity,
\begin{equation}
\sla{p} \gamma^0 \pm m \gamma^0 = \sqrt{2} {\cal P}_+ p^+ + \sqrt{2} {\cal P}_-
p^- \pm {m \over \sqrt{2}} (\gamma^+ + \gamma^-) \; ,
\end{equation}
implies the following projections,
\begin{equation}
{\cal P}_+ (\sla{p} \gamma^0 \pm m \gamma^0) {\cal P}_+ = \sqrt{2} p^+ {\cal 
P}_+ \;\; , \;\; {\cal P}_+(\sla{p} \gamma^0 \pm m \gamma^0) {\cal P}_- 
\gamma^+ = \pm \sqrt{2} m {\cal P}_+ \; ,
\ee
\be
\gamma^- {\cal P}_-(\sla{p} \gamma^0 \pm m \gamma^0) {\cal P}_+ = \pm \sqrt{2}
m {\cal P}_+ \;\; , \;\; {\cal P}_- (\sla{p} \gamma^0 \pm m \gamma^0) 
{\cal P}_- = \sqrt{2} p^- {\cal P}_- \; . \label{identities1}
\ee
As with the particle production probability, we can drop the $+-$ and $-+$
correlators in the large $L$ limit. The $++$ and $--$ correlators are,
\begin{eqnarray}
\lefteqn{\Bigl\langle \Omega \Bigl\vert \psi_+^\dagger(0,w) \psi_+(0,v) - {\rm 
Tr}\Bigl[ \psi_+(0,v) \psi_+^\dagger(0,w) \Bigr] \Bigr\vert \Omega \Bigr\rangle
} \nonumber \\
& & \hspace{2cm} = \frac1{\sqrt{2}} \Biggl\{\int_{-\infty}^0 {dp^+ \over 2\pi} 
- \int_0^\infty {dp^+ \over 2 \pi} \Biggr\} e^{-i p^+ (v-w)} \; , \;
\end{eqnarray}
\begin{equation}
\Bigl\langle \Omega \Bigl\vert \psi_-^\dagger(y,-L) \psi_-(u,-L) - {\rm 
Tr}\Bigl[\psi_-(u,-L) \psi_-^\dagger(y,-L) \Bigr] \Bigr\vert \Omega\Bigr\rangle
= {i \over {\sqrt{2} \pi}} {\cal P}\Bigl({1 \over {u-y}}\Bigr) \; .
\end{equation}

The reduction strategy is quite similar to that used in Section 4 for the
probability of particle production. First perform the integrals over $v$ and 
$w$,
\be
\int_{-L}^\infty dv e^{i(k^++i/L)v} e^{-il^+(v+L)} = {{ie^{-i(k^++i/L)L}} \over
{k^+-l^++i/L}} \; .
\ee
Next make the appropriate change of change variables from $k^+$ and $q^+$ to,
\be
a\equiv (k^+-l^+)L \;\;\; {\rm and}\;\;\; b\equiv (q^+-l^+)L\; \; ,
\ee
and take the $L \rightarrow \infty$ limit of the mode functions ${\cal E}$ 
using,
\begin{eqnarray}
\lefteqn{\lim_{L \rightarrow \infty} {\cal E}\Bigl(0,x^+;p^+ + \frac{a}{L}
\Bigr)\; {\cal E}^*\Bigl(0,x^+;p^+ + \frac{b}{L}\Bigr) } \nonumber \\
& & \hspace{3cm} = \exp\Bigl[-2 \pi \lambda(p^+) \theta(p^+) \theta\Bigl(e 
A_-(x^+) - p^+\Bigr)\Bigr] , \;
\end{eqnarray}
\begin{eqnarray}
\lefteqn{\lim_{L \rightarrow \infty} {\cal E}\Bigl(X(r^+),x^+;r^+ + \frac{a}{L}
\Bigr) \; {\cal E}^*\Bigl(X\Bigl(r^+ - \frac{c}{L}\Bigr),x^+;r^+ + 
\frac{b-c}{L}\Bigr) } \nonumber \\
& & \hspace{3cm} = \exp\Bigl[\lambda(r^+) \Bigl(-2 \pi - i \ln(a+i) + i 
\ln(b-i) \Bigr)\Bigr] \; .
\end{eqnarray}
(We have assumed $r^+ > 0$ in taking the last limit.) Finally the $a$ and $b$ 
integrations are performed using (\ref{aint}), and the result is simplified 
with the identity (\ref{Loba}) of Lobachevskiy.

Applying this procedure to the $++$ term gives,
\begin{eqnarray}
\lefteqn{J^+_{(++)} = {e \over 2} e^{ieA_-(x^+)\Delta} \Biggl\{\int_{-\infty}^0
{{dp^+} \over {2\pi}} - \int_0^\infty {{dp^+} \over {2\pi}}\Biggr\} \int_{-
\infty}^\infty {{dk^+} \over {2\pi}} {{e^{-i(k^++i/L) (L + x^- + \Delta)}}
\over {k^+-p^++i/L}} } \nonumber \\
& & \hspace{2.5cm} \times {\cal E}(0,x^+;k^+) \int_{-\infty}^\infty {{dq^+} 
\over {2\pi}} {{e^{i(q^+-i/L)(L+x^-)}} \over {q^+-p^+-i/L}} {\cal E}^*(0,x^+;
q^+)  \; , \\
& & = {e \over 2} \Biggl\{\int_{-\infty}^0 {{dp^+} \over {2\pi}} - 
\int_0^\infty {{dp^+} \over {2\pi}}\Biggr\} e^{-i(p^+-eA_-(x^+)) \Delta}
\int_{-\infty}^\infty {{da} \over {2\pi}} {{e^{-i(a+i) (1+(x^- + \Delta)/L)}}
\over {a+i}} \nonumber \\
& & \hspace{1.5cm} \times {\cal E}\Bigl(0,x^+;p^+ + \frac{a}{L}\Bigr) 
\int_{-\infty}^\infty {{db} \over {2\pi}} {{e^{i(b-i)(1+x^-/L)}} \over {b-i}}
{\cal E}^*\Bigl(0,x^+;p^+ + \frac{b}{L}\Bigr) , \quad \\ 
& & \longrightarrow {e\over 2} \Biggl\{\int_{-\infty}^0 - \int_{eA_-(x^+)}^{
\infty} - \int_0^{eA_-(x^+)} e^{-2\pi\lambda(p^+)}\Biggr\} {dp^+ \over 2 \pi}
e^{-i[p^+-eA_-(x^+)]\Delta} \; , \\
& & = {e \over 2} \Biggl\{\int_{-\infty}^{eA_-(x^+)} - \int_{eA_-(x^+)}^\infty 
- \int_0^{eA_-} \Bigl[1 + e^{-2 \pi \lambda(p^+)}\Bigr] \Biggr\} {{dp^+} 
\over {2\pi}} e^{-i[p^+-eA_-] \Delta} \; , \\
& & = {e \over 2} \Biggl\{ {i \over {\pi \Delta}} - \int_0^{eA_-(x^+)} {{dp^+}
\over {2\pi}} \Bigl[1 + e^{-2\pi\lambda(p^+)}\Bigr] e^{-i [p^+-eA_-(x^+)] 
\Delta} \Biggr\} \; . \label{J+1}
\end{eqnarray}

We begin the reduction of the $--$ term by changing variables from $u$ and $y$
to $r^+ \equiv e A_-(u)$ and $s^+ \equiv e A_-(y)$,
\begin{eqnarray}
\lefteqn{J^+_{(--)} = {e \over {m^2}} e^{ieA_-(x^+) \Delta} \int_0^{eA_-(x^+)}
dr^+ \lambda(r^+) \int_0^{eA_-(x^+)} ds^+ \lambda(s^+) } \nonumber \\
& & \times {i \over \pi} {\cal P}\Biggl({1 \over X(r^+)-X(s^+)}\Biggr) 
\int_{-\infty}^\infty {{dk^+} \over {2\pi}} {{e^{-i(k^+ + i/L) (L + x^- + 
\Delta)}} \over {k^+-r^++i/L}} {\cal E}\Bigl(X(r^+),x^+;k^+\Bigr) \nonumber \\
& & \hspace{3cm} \times \int_{-\infty}^\infty {{dq^+} \over {2\pi}} {{e^{i(q^+
- i/L) (L + x^-)}} \over {q^+-s^+-i/L}} {\cal E}^*\Bigl(X(s^+),x^+;q^+\Bigr) 
\; , \\
& & = {i e \over \pi m^2} \int_0^{eA_-(x^+)} dr^+ \lambda(r^+) e^{-i[r^+ -
eA_-(x^+)] \Delta} \int_0^{eA_-(x^+)} ds^+ \lambda(s^+) \nonumber \\
& & \times {\cal P}\Biggl( { e^{-i(r^+-s^+)(L+x^-)} \over X(r^+) - X(s^+)}
\Biggr) \int_{-\infty}^\infty {{da} \over {2\pi}} {{e^{-i(a+i) [1 + (x^- + 
\Delta)/L]}} \over {a+i}} {\cal E}\Bigl(X(r^+),x^+;r^+ + \frac{a}{L}\Bigr) 
\nonumber \\
&& \hspace{3cm} \times \int_{-\infty}^\infty {{db} \over {2\pi}} {{e^{i(b-i)
(1 + x^-/L)}} \over {b-i}} {\cal E}^*\Bigl(X(s^+),x^+;s^+ + \frac{b}{L}\Bigr) 
\; .
\end{eqnarray}
Now change variables from $s^+$ to $c\equiv (r^+-s^+)L$ and expand,
\begin{equation}
X(s^+) = X(r^+ - c/L) = X(r^+) - {{2 \lambda(r^+)} \over {m^2}} {c \over L} + 
{\cal O}(L^{-2}) \; .
\end{equation}
The large $L$ limit is therefore,
\begin{eqnarray}
\lefteqn{J^+_{(--)} \longrightarrow {e\over 2} \int_0^{eA_-(x^+)} dr^+ \lambda(
r^+) e^{-i[r^+-eA_-(x^+)] \Delta} e^{-2\pi \lambda(r^+)} \int_{-\infty}^\infty 
{dc \over \pi} {{\sin{c}} \over c} } \nonumber \\
& & \hspace{1cm} \times \int_{-\infty}^\infty {{da} \over{2\pi}} {{e^{-i(a+i)}}
\over {a+i}} e^{-i\lambda(r^+) \ln(a+i)} \int_{-\infty}^\infty {{db} \over 
{2\pi}} {{e^{i(b-i)}} \over {b-i}} e^{i\lambda(r^+) \ln(b-i)} , \\
& & = {e \over 2} \int_0^{eA_-(x^+)} dr^+ \lambda(r^+) e^{-i[r^+ - eA_-(x^+)]
\Delta} 
{e^{-\pi \lambda(r^+)} \over \Vert \Gamma(1+i\lambda(r^+)) \Vert^2} \; , \\
& & = {e \over 2} \int_0^{eA_-(x^+)} dr^+ \lambda(r^+) e^{-i[(r^+ - eA_-(x^+)]
\Delta} \Biggl[{1 - e^{-2\pi\lambda(r^+)} \over 2 \pi \lambda(r^+)} \Biggr]
\; , \\
& & = {e\over 2} \int_0^{eA_-(x^+)} {{dr^+} \over {2\pi}} e^{-i [(r^+ - 
eA_-(x^+)] \Delta} \Bigl[1 - e^{-2 \pi \lambda(r^+)}\Bigr] \; .
\end{eqnarray}

Adding the $++$ and $--$ contributions gives,
\begin{eqnarray}
\lefteqn{\lim_{L\rightarrow \infty} \Bigl\langle \Omega \Bigl\vert 
J^+(x^+;x^- + \Delta,x^-) \Bigr\vert \Omega \Bigr\rangle = {e \over 2} 
\Biggl\{{i \over {\pi \Delta}} } \nonumber \\
& & \hspace{3cm} - 2 \int_0^{eA_-(x^+)} {{dp^+} \over {2 \pi}} e^{-2 \pi 
\lambda(p^+)} e^{-i [p^+ - e A_-(x^+)] \Delta} \Biggr\} . \label{J+delta}
\end{eqnarray}
The linear divergence vanishes upon Hermitization, so we can take the splitting
parameter $\Delta$ to zero,
\begin{eqnarray}
\lefteqn{\lim_{L \rightarrow \infty} \Bigl\langle \Omega \Bigl\vert 
J^+_S(x^+;x^- + \Delta,x^-) \Bigr\vert \Omega \Bigr\rangle} \nonumber \\
& = & -e \int_0^{eA_-(x^+)} {{dp^+} \over {2\pi}} e^{-2\pi\lambda(p^+)} 
\cos\Bigl[\Bigl(p^+- eA_-(x^+)\Bigr) \Delta\Bigr] \; , \\
& \longrightarrow & - e \int_0^{eA_-(x^+)} {{dp^+} \over {2 \pi}} e^{-2 \pi
\lambda(p^+)} \; . \label{finalJ+}
\end{eqnarray}

The interpretation of this result is straightforward. $J^+$ is the light-cone
charge density, so we expect it to grow as more and more positrons are created.
(Recall that the electrons immediately leave the light-cone manifold as 
depicted in Fig.~2.) Each positron carries charge $-e$; the probability for
mode $p^+$ to be created is $\exp[-2 \pi \lambda(p^+)]$ (from equation 
(\ref{probability})); and the number of modes per unit volume is $dp^+/{2\pi}$.
We therefore expect the increment to obey,
\begin{equation}
dJ^+ = \Bigl( -e \Bigr) \times \Bigl(e^{-2\pi \lambda(p^+)} \Bigr) \times
\Bigl({dp^+ \over 2\pi}\Bigr) \; . \label{dJ+}
\end{equation}
For monotonically increasing $e A_-(x^+)$, all the modes in the interval $0 <
p^+ < e A_-(x^+)$ will have contributed, which gives precisely (\ref{finalJ+}).
This is a powerful check on the fundamental correctness of our formalism as 
well as on our proper application of it.

It remains to compute the expectation value of $J^-$. First note that we can
express $J^-(x^+,y^+;x^-)$ in terms of $\psi_+$ using the field equation 
$\psi_-=(i/m)\gamma^+ i\partial_+ \psi_+$,
\begin{eqnarray}
\lefteqn{J^-(x^+,y^+;x^-) = {\sqrt{2} e \over m^2} {\partial\over \partial x^+}
{\partial \over {\partial y^+}} \Biggl\{\psi_+^\dagger(y^+,x^-) 
\psi_+(x^+,x^-) } \nonumber \\
& & \hspace{6cm} - {\rm Tr}\Bigl[ \psi_+(x^+,x^-) \psi_+^\dagger(y^+,x^-) 
\Bigr]\Biggr\} . \quad
\end{eqnarray}
As before we can use the general solution (\ref{psi+}) to reduce the 
expectation value of this operator to a sum of four correlators on the initial
value surface. Also as before, the $+-$ and $-+$ correlators vanish for 
infinite $L$. We therefore require only,
\begin{eqnarray}
\lefteqn{J^-_{(++)} \equiv {e \over {m^2}} {\partial \over {\partial x^+}}
{\partial \over {\partial y^+}} \int_{-\infty}^\infty {dk^+ \over 2\pi}
e^{-i(k^+ + i/L)(x^-+L)} {\cal E}(0,x^+;k^+) } \nonumber \\
& & \times \int_{-\infty}^\infty {{dq^+} \over {2\pi}} e^{i(q^+ - i/L)(x^-+L)} 
{\cal E}^*(0,y^+;q^+) \nonumber \\
& & \hspace{1cm} \times \Biggl\{\int_{-\infty}^0 {dp^+ \over 2\pi} - \int_0^{
\infty} {dp^+ \over 2\pi}\Biggr\} {1 \over {(k^+ - p^+ + i/L) (q^+ - p^+ - 
i/L)}} , \quad
\end{eqnarray}
and,
\begin{eqnarray}
\lefteqn{J^-_{(--)} \equiv {i e \over 2 \pi} {\partial \over {\partial x^+}}
{\partial \over {\partial y^+}} \int_{-\infty}^\infty {{dk^+} \over {2\pi}}
e^{-i(k^+ + i/L)(x^-+L)} \int_{-\infty}^\infty {{dq^+} \over {2\pi}} e^{i(q^+
- i/L)(x^-+L)} } \nonumber \\
& & \times \int_0^{x^+} {du \; {\cal E}(u,x^+;k^+) \over {k^+ - eA_-(u) + i/L}}
\int_0^{y^+} {dz \; {\cal E}^*(z,y^+;q^+) \over {q^+ - eA_-(z) - i/L}}
{\cal P}\Biggl({1 \over {u-z}}\Biggr) \; . \label{<J->3}
\end{eqnarray}

After taking the $x^+$ and $y^+$ derivatives, and performing the $p^+$ 
integrals, the $++$ term becomes,
\begin{eqnarray}
\lefteqn{J^-_{(++)} = - {{i e m^2} \over 4 \pi} \int_{-\infty}^\infty {{dk^+} 
\over {2 \pi}} {{e^{-i(k^+ + i/L)(x^- + L)}} \over {k^+ - eA_-(x^+) + i/L}} 
{\cal E}(0,x^+;k^+) \int_{-\infty}^{\infty} {dq^+ \over 2 \pi} } \nonumber \\
& & \times {e^{i(q^+ - i/L)(x^- + L)} \over q^+ - eA_-(y^+) - i/L} 
{{\cal E}^*(0,y^+;q^+) \over {k^+ - q^+ + 2 i/L}} \Biggl[\pi + i 
\ln\Biggl({{k^++i/L} \over {q^+-i/L}}\Biggr) \Biggr] \; . \label{J-1}
\end{eqnarray}
Acting the $x^+$ derivative on $J^-_{(--)}$ produces two terms, one where the 
derivative hits the upper limit of the $u$ integration and the other where it 
hits the mode function ${\cal E}(u,x^+;k^+)$. The first of these terms simply 
sets $u = x^+$, which makes the mode function unity. The second term brings 
down a factor of $-m^2/2$ divided by $[k^+ - e A_-(x^+) + i/L]$. Acting the 
$y^+$ derivative gives two similar terms, with the result that $J^-_{(--)}$ 
can be written as the sum of the following four expressions,
\be
J^-_{(a)} = {{ie} \over {2\pi}}{{e^{-ie [A_-(x^+)-A_-(y^+)] (x^-+L)}} \over
{x^+ - y^+}} \; , \label{J-4a}
\ee
\begin{eqnarray}
\lefteqn{J^-_{(b)} = {i e m^2 \over 4\pi}\int_0^{y^+} {\cal P}\Bigl({dz \over 
x^+-z} \Bigr) } \nonumber \\
& & \hspace{1cm} \times \int_{-\infty}^\infty {{dq^+} \over {2\pi}} {{e^{i[q^+ 
- eA_-(x^+) -i/L] (x^-+L)} {\cal E}^*(z,y^+;q^+)} \over {[q^+- eA_-(z) - i/L] 
[q^+ - eA_-(y^+) -i/L]}} \; , \label{J-4b}
\end{eqnarray}
\begin{eqnarray}
\lefteqn{J^-_{(c)} = {i e m^2 \over 4\pi} \int_0^{x^+} {\cal P}\Bigl({du \over 
{u - y^+}}\Bigr) } \nonumber \\
& & \hspace{1cm} \times \int_{-\infty}^\infty {{dk^+}\over{2\pi}} {{e^{-i[k^+ - 
eA_-(y^+) + i/L] (x^-+L)} {\cal E}(u,x^+;k^+)} \over {[k^+ - eA_-(u) + i/L]
[k^+ - eA_-(x^+) + i/L]}} \; , \label{J-4c}
\end{eqnarray}
\begin{eqnarray}
\lefteqn{ J^-_{(d)} = {{i e m^4} \over 8 \pi} \int_0^{x^+} du \int_0^{y^+} dz 
{\cal P}\Bigl({1 \over {u-z}}\Bigr) } \nonumber \\
& & \hspace{.5cm} \times \int_{-\infty}^\infty {{dk^+} \over {2\pi}} {{e^{-i
(k^+ + i/L)(x^-+L)} {\cal E}(u,x^+;k^+)} \over {[k^+ - eA_-(u) + i/L] [k^+ - 
eA_-(x^+) + i/L]}} \nonumber \\
&& \hspace{1cm} \times \int_{-\infty}^\infty {{dq^+} \over {2\pi}} {{e^{i(q^+
- i/L) (x^-+L)} {\cal E}^*(z,y^+;q^+)} \over {[q^+ - eA_-(z) - i/L] [q^+ - 
eA_-(y^+) - i/L]}} \; . \label{J-4d}
\end{eqnarray}

At this point the reduction of $\langle J^- \rangle$ deviates somewhat from the
procedure we used for $\langle J^+ \rangle$, and it is well to comment on the 
reasons for this before proceeding. First note that the only ultraviolet 
divergent contribution comes from the single term --- $J^-{(a)}$ --- for which
we can obtain an explicit result in terms of elementary functions before taking
the large $L$ limit. All the other terms remain finite as we take $y^+ 
\longrightarrow x^+$ at fixed $L$. Second, note that all the other terms vanish
at $x^- = -L$ because we can close the $k^+$ and $q^+$ contours above and 
below, respectively, where the integrand in analytic. The physical reason for
this is that $x^- = -L$ is the initial value surface upon which our state
$\vert \Omega \rangle$ agrees with the $A_- = 0$ vacuum that has zero 
current.\footnote{At $x^- = -L$ the term $J^-_{(a)}$ degenerates to a pure
imaginary, linear divergence which vanishes upon Hermitization.} 

A third important observation is that all the other terms diverge linearly when
the large $L$ limit is taken at finite $x^-$ after setting $y^+ = x^+$. The 
physical reason for this is that, just as $J^+$ is the light-cone charge 
density, so $J^-$ gives the light-cone charge {\it flux}. Although the 
electrons make no contribution to $J^+$ because they immediately leave the 
manifold moving parallel to the $x^-$ axis, they {\it do} contribute to $J^-$ 
for the very same reason. The electron current flowing through any fixed value 
of $x^-$ consists of the flux originating in each element $dx^-$, all the way 
back to $x^- = -L$. In analogy with (\ref{dJ+}) we expect the increment from 
each volume element $dx^-$ to be the electron charge times the probability for 
creation times the rate at which modes pass through the critical point $p^+ = 
e A_-(x^+)$,
\begin{equation}
dJ^- = \Bigl( +e \Bigr) \times \Bigl(e^{-2\pi \lambda(e A_-(x^+))} \Bigr) 
\times \Bigl({e A_-^{\prime}(x^+) dx^- \over 2\pi}\Bigr) \; . \label{dJ-}
\end{equation}
Since this expression does not depend upon $x^-$, integrating it from $-L$ to
$x^-$ adds a factor of $(L + x^-)$, which is the origin of the linear 
divergence in the large $L$ limit.

This last observation implies that we cannot take the large $L$ limit in
evaluating the expectation value of $J^-$. However, the fact that we know the
value of $J^-$ at $x^- = - L$ (zero) suggests that we can equally well compute
the expectation value of $\partial_- J^-$ --- which {\it does} have a finite 
large $L$ limit --- and then integrate to obtain the undifferentiated current.
That is the strategy we shall follow. To simplify the computation we shall also
take the coincidence limit before letting $L$ go to infinity.

From (\ref{J-1}) we obtain,
\begin{eqnarray}
\lefteqn{\partial_- J^-_{(++)} = {i e m^2 \over 4\pi} \int_{-\infty}^\infty 
{{dk^+} \over {2\pi}} {{e^{-i(k^+ + i/L)(x^-+L)}} \over {k^+ - eA_-(x^+) +i/L}}
{\cal E}(0,x^+;k^+) } \nonumber \\
& & \times \int_{-\infty}^{\infty} {dq^+ \over 2\pi}
{e^{i(q^+ - i/L)(x^-+L)} {\cal E}^*(0,y^+;q^+) \over {q^+ - eA_-(y^+) - i/L}} 
\Biggl[i \pi - \ln\Biggl({k^+ + i/L \over {q^+ - i/L}}\Biggr)\Biggr] . \quad
\end{eqnarray}
Making the change of variables $a \equiv [k^+ - eA_-(x^+)] L$ and $b \equiv 
[q^+ - eA_-(y^+) ]L$, and taking the limits $x^+\to y^+$ and $L\to\infty$
gives,
\begin{eqnarray}
\lefteqn{\partial_- J^-_{(++)}\longrightarrow  -{{e m^2} \over 4} \int_{-\infty
}^\infty {{da} \over {2\pi}} {{e^{-i(a+i)}} \over {a+i}} e^{i\lambda(eA_-(x^+))
\ln(a+i)} } \nonumber \\
& & \hspace{3cm} \times \int_{-\infty}^\infty {{db} \over {2\pi}} {{e^{i(b-i)}}
\over {b-i}} e^{-i \lambda(eA_-(x^+)) \ln(b-i)} \; , \\
& & = -{{e^2 A'_-(x^+)} \over {4\pi}} \Bigl[1 - e^{-2 \pi \lambda(eA_-(x^+))}
\Bigr] \; . \label{d-J-1}
\end{eqnarray}

To evaluate the $--$ terms we first note that the coincidence limit of 
$\partial_- J^-_{(a)}$ is real and finite,
\be
\partial_- J^-_{(4a)} \to {{e^2 A'_-(x^+)}\over{2\pi}} \; . \label{d-J-4a}
\ee
Next combine the coincidence limits of the $b$, $c$ and $d$ terms to reach,
\begin{eqnarray}
\lefteqn{\lim_{y^+ \rightarrow x^+} \partial_- J^-_{(b-d)} = - {{i e m^2} \over
4 \pi} {\partial \over {\partial x^+}} \int_0^{x^+} du \int_0^{x^+} {\cal P}
\Bigl({dz \over {u-z}}\Bigr) \int_{-\infty}^\infty {dk^+ \over 2\pi} 
{\cal E}(u,x^+;k^+) } \nonumber \\
& & \hspace{.5cm} \times {e^{-i(k^+ + i/L)(x^- + L)} \over {k^+ - eA_-(u) + 
i/L}} \int_{-\infty}^\infty {dq^+ \over 2\pi} {e^{i(q^+ - i/L)(x^- + L)} {\cal 
E}^*(z,x^+;q^+) \over {q^+ - eA_-(z) - i/L}} . \qquad \label{d-J-4d}
\end{eqnarray}
Now change variables from $k^+$, $q^+$ and $z$ to,
\begin{equation}
a \equiv [k^+-eA_-(u)] L \quad , \quad b \equiv [q^+ - eA_-(z)] L \quad , \quad
c \equiv e [A_-(u)-A_-(z)] L \; ,
\end{equation}
and take the large $L$ limit,
\begin{eqnarray}
\lefteqn{\lim_{L \rightarrow \infty} \lim_{y^+ \rightarrow x^+} \partial_- 
J^-_{(b-d)} = -{{e m^2} \over 4} {\partial \over {\partial x^+}} \int_0^{x^+} 
du e^{-2 \pi \lambda(e A_-(u))} \int_{-\infty}^{\infty} dc {\sin c \over c} } 
\nonumber \\
& & \times \int_{-\infty}^\infty {{da} \over{2\pi}} {{e^{-i(a+i)}} \over 
{a+i}} e^{-i \lambda(e A_-(u)) \ln(a+i)} \int_{-\infty}^\infty {{db} \over 
{2\pi}} {{e^{i(b-i)}} \over {b-i}} e^{i \lambda(e A_-(u)) \ln(b-i)} , \quad
\\ \label{d-J-4d2}
& & = -{{e^2 A'_-(x^+)} \over {4\pi}} \Bigl[1 - e^{-2 \pi \lambda(eA_-(x^+))}
\Bigr] \; .
\end{eqnarray}

Combining with (\ref{d-J-1}) and (\ref{d-J-4a}) we obtain, 
\be
\lim_{L \rightarrow \infty} \lim_{y^+ \rightarrow x^+} \partial_- \Bigl\langle
\Omega \Bigl\vert J^-(x^+,y^+;x^-) \Bigr\vert \Omega \Bigr\rangle = {{e^2 
A'_-(x^+)} \over {2\pi}} e^{-2\pi\lambda(eA_-(x^+))} \; . \label{d-J-}
\ee
Since this is already real, Hermitization makes no change. The fact that these
explicit calculations are in complete agreement with our physics-based 
expectation (\ref{dJ-}) is another impressive check.

Our final results for the vector current are,
\begin{eqnarray}
\lim_{y^- \rightarrow x^-} \Bigl\langle \Omega \Bigl\vert J^+_S(x^+;x^-,y^-) 
\Bigr\vert \Omega \Bigr\rangle & = & -e \int_0^{e A_-(x^+)} {dp^+ \over 2\pi} 
e^{-2 \pi \lambda(p^+)} + \dots \; , \label{J+final} \\
\lim_{y^- \rightarrow x^-} \Bigl\langle \Omega \Bigl\vert J^-_S(x^+,y^+;x^-) 
\Bigr\vert \Omega \Bigr\rangle & = & {{e^2 A'_-} \over {2\pi}} e^{-2 \pi
\lambda(eA_-)} \Bigl(L + x^-\Bigr) + \dots \label{J-final}
\end{eqnarray}
Here the dots indicate terms which vanish as $L$ goes to infinity. Note that
the divergence of the vector current vanishes, as it should.

\section{The axial vector anomaly}

It is convenient to point split the pseudoscalar in both directions,
\begin{eqnarray}
\lefteqn{
J_5(x^+,y^+;x^-,y^-)\equiv {1 \over \sqrt{8}} \exp\Biggl[i e (x^--y^-)
\int_0^1 d\tau A_-\Bigl(y^+ + \tau (x^+ - y^+) \Bigr)\Biggr] } \nonumber \\
& & \times \Biggl\{\psi_-^\dagger(y^+,y^-) \gamma^+ \psi_+(x^+,x^-) -
\psi_+^\dagger(y^+,y^-) \gamma^- \psi_-(x^+,x^-) \nonumber \\
& & - {\rm Tr}\Bigl[\psi_+(x^+,x^-) \psi_-^\dagger(y^+,y^-) \gamma^+\Bigr] + 
{\rm Tr}\Bigl[\psi_-(x^+,x^-) \psi_+^\dagger(y^+,y^-) \gamma^- \Bigr]\Biggr\} .
\quad \qquad
\end{eqnarray}
This can be rewritten in terms of $\psi_+$ alone by using (\ref{eqn2}),
\begin{eqnarray}
\lefteqn{J_5(x^+,y^+;x^-,y^-) = {-i \over \sqrt{2} m} \exp\Biggl[ i e (x^- - 
y^-) \int_0^1 d\tau A_-\Bigl(y^+ + \tau (x^+-y^+) \Bigr)\Biggr] } \nonumber \\
& & \quad \times \Biggl({\partial \over {\partial x^+}} + {\partial \over 
{\partial y^+}} \Biggr) \Biggl\{\psi_+^\dagger(y^+,y^-) \psi_+(x^+,x^-) 
\nonumber \\
& & \hspace{3cm} - {\rm Tr}\Bigl[(\psi_+(x^+,x^-) \psi_+^\dagger(y^+,y^-) 
\Bigr]\Biggr\} \qquad \; , \\
& & = {-i \over \sqrt{2} m} \Biggl({\partial \over {\partial x^+}} + {\partial 
\over {\partial y^+}} - ie {{A_-(x^+) - A_-(y^+)} \over {x^+-y^+}} (x^--y^-) 
\Biggr) \nonumber \\
& & \hspace{.5cm} \times \exp\Biggl[ie (x^- - y^-) \int_0^1 d\tau A_-\Bigl(y^+ +
\tau (x^+ - y^+) \Bigr)\Biggr] \nonumber \\
& & \hspace{1cm} \times \Biggl\{\psi_+^\dagger(y^+,y^-) \psi_+(x^+,x^-) - {\rm
Tr}\Bigl[\psi_+(x^+,x^-) \psi_+^\dagger(y^+,y^-)\Bigr] \Biggr\} \; . \quad
\end{eqnarray}
Now take the $+$ coordinates to coincidence,
\begin{eqnarray}
\lefteqn{J_5(x^+,x^+;x^-,y^-) = {-i \over e m} {\partial \over {\partial x^+}} 
J^+(x^+;x^-,y^-)} \nonumber \\
& & \hspace{2cm}  - {1 \over m} A'_-(x^+) (x^- - y^-) J^+(x^+;x^-,y^-) \; . 
\label{J5J+}
\end{eqnarray}
This strong operator equation is still well regulated by the point splitting
of the $-$ coordinates.

In the large $L$ limit the expectation value of (\ref{J5J+}) gives,
\begin{eqnarray}
\lefteqn{\lim_{L \rightarrow \infty} \Bigl\langle \Omega \Bigl\vert J_5(x^+,x^+
;x^-,y^-) \Bigr\vert \Omega \Bigr\rangle} \nonumber \\
& & = {-i \over e m} \Biggl({\partial \over {\partial x^+}} -i e A'_-(x^+) (x^-
- y^-) \Biggr) \lim_{L \rightarrow \infty} \Bigl\langle \Omega \Bigl\vert 
J^+(x^+;x^-,y^-) \Bigr\vert \Omega \Bigr\rangle . \quad
\end{eqnarray}
We computed the expectation value on the right hand side in (\ref{J+delta}).
Substituting this relation and taking the $-$ points to coincidence gives,
\begin{eqnarray}
\lefteqn{ \lim_{y^- \rightarrow x^-} \lim_{L \rightarrow \infty} \Bigl\langle 
\Omega \Bigl\vert J_5(x^+,x^+;x^-,y^-) \Bigr\vert \Omega \Bigr\rangle =}
\nonumber \\
& & -{i \over 2 m} \lim_{y^- \rightarrow x^-} \Biggl({\partial \over \partial 
x^+} -i e A'_-(x^+) (x^- - y^-) \Biggr) \Biggl\{{i \over {\pi (x^- - y^-)}}
\nonumber \\
& & \hspace{2cm} - 2 \int_0^{eA_-(x^+)} {{dp^+} \over {2 \pi}} e^{-2 \pi 
\lambda(p^+)} e^{-i [p^+ - e A_-(x^+)] (x^- - y^-)} \Biggr\} \; , \\
& & = {-i e \over 2\pi m} A'_-(x^+) \Bigl[1 - e^{-2 \pi \lambda(e A_-(x^+))}
\Bigr] \; .
\end{eqnarray}
The axial vector anomaly is the deviation from the naive divergence equation 
(\ref{naive2}), 
\begin{equation}
\lim_{y^- \rightarrow x^-} \lim_{L \rightarrow \infty} \Bigl\langle \Omega 
\Bigl\vert 2 \partial_+ J^+(x^+;x^-,y^-) - 2 i e m J_5(x^+,x^+;x^-,y^-) 
\Bigr\vert \Omega \Bigr\rangle = {e^2 \over \pi} E(x^+) \; .
\end{equation}
That we get it exactly right is yet another check. That it does {\it not} come 
out right when one neglects the $\psi_-$ initial value data at $x^- = -L$ is an
additional illustration of the essential role these terms play.

\section{Discussion}

This work was undertaken to exploit a crucial extension of our earlier solution 
\cite{ttw} for the Dirac operator in the presence of an electric field which 
can depend arbitrarily upon the light-cone time parameter $x^+$. To properly
resolve the ambiguity at $p^+ = 0$ we had discovered that it is essential to
specify $\psi_-(x^+,-L)$ for $x^+ > 0$ in addition to $\psi_+(0,x^-)$ for $x^-
> -L$. (See Fig.~1.) In our previous solution it was necessary to take $L$ to 
infinity, {\it at the operator level}, before computing expectation values. 
Needless to say, this could only be done distributionally, with the inevitable 
restriction that the limiting form of the operator not be multiplied by any
other operator which can behave badly in the large $L$ limit. The result was
that we could handle $J^+$, but not $J^-$ or $J_5$. Our specification of the
vacuum was also cumbersome and not obviously in agreement with the known 
massless limit in $1+1$ dimensions.

Our new solution (\ref{psi+},\ref{psi-}) is exact for any $L$, and it can be
employed in any operator product with only the usual regularization. We also
have a transparent definition of the vacuum as the state which agrees with the
$E = 0$ vacuum on the initial value surface. So one computes the expectation
value of any operator at $(x^+,x^-)$ by first using the solution 
(\ref{psi+},{\ref{psi-}) to express the VEV in terms of correlators of the
initial value operators. Then one computes these correlators by using free 
Dirac theory with zero electric field. The result is a vast expansion of the
things we can do. In this paper we computed the probability for pair 
production, as well as the one loop expectation values of the vector and axial 
vector currents and of the pseudoscalar $J_5$. All our results are valid for
any mass, and for any positive electric field, under the assumption that both
are independent of $L$.

It is especially significant that we recover the well known result for the 
axial vector anomaly, for the first time ever in massive QED on the light-cone.
The obstacle in previous efforts to achieve this seems to have been the failure
to properly resolve the ambiguity at $p^+ = 0$ by specifying $\psi_-(x^+,-L)$. 
In this we were fortunate that the background's peculiar propensity to pull 
{\it each} positive $k^+$ through the singularity at $p^+(x^+) \equiv k^+ - 
e A_-(x^+) = 0$ forced us to come to grips with the problem that remains at
$k^+ = 0$ for zero electric field. Our work provides 
an explicit contradiction to the belief, that it is 
consistent to use data on only $x^+=0$, provided one imposes appropriate 
boundary conditions on the second characteristic $x^-=$constant. This was shown
to be true for free field theory \cite{heinzl}, but does not seem to be 
valid in the presence of spacetime dependent backgrounds, or in more general 
interacting theories. There is no mixing between modes in the free theory, 
so making the $p^+=0$ mode nondynamical does not affect the other modes. 
With a positive electric field, more and more of the $k^+ > 0$ modes are 
pulled through the singularity, after which they must incorporate operators 
from the $x^- = -L$ surface if the canonical anti-commutation relations are 
to be preserved. Note that this remains true even in the limit of infinite 
$L$. Without the $x^- =-L$ operators one finds,
\begin{eqnarray}
\lefteqn{ \lim_{L \rightarrow \infty} \Bigl\{\psi_+(x^+,x^-),\psi^{
\dagger}_+(x^+,y^-) \Bigr\} } \nonumber \\
& & = \frac{{\cal P}_+}{\sqrt{2}} \Biggl(\delta(x^- - y^-) - 
\int_0^{eA_-} {dk^+ \over 2\pi} \Bigl[1 - e^{-2\pi \lambda(k^+)}\Bigr] 
e^{-i k^+ (x^- - y^-)} \Biggr) \; . \quad
\end{eqnarray}

We worked in $1+1$ dimensions because it is simple to do so, but the exact 
solution generalizes to any spacetime dimension $D$. Let us represent the 
$(D-2)$ transverse coordinates with a tilde thusly, $\widetilde{x}$. Without
overburdening the notation too much we can also employ a tilde to denote the 
transverse Fourier transform of the dynamical variables,
\begin{equation}
\widetilde{\psi}_{\pm}(x^+,x^-,\widetilde{k}) \equiv \int d^{D-2}\widetilde{x}
e^{-i \widetilde{k} \cdot \widetilde{x}} \psi_{\pm}(x^+,x^-,\widetilde{x}) \; .
\end{equation}
The higher dimensional solution is,
\begin{eqnarray}
\lefteqn{\widetilde{\psi}_+(x^+,x^-,\widetilde{k}) = \int_{-L}^{\infty}dv 
\int_{-\infty}^{+\infty} {{dk^+}\over{2\pi}} e^{i(k^++i/L) (v - x^-)} 
\Biggl\{{\cal E}(0,x^+;k^+,\widetilde{k}) \widetilde{\psi}_+(0,v,
\widetilde{k})} \nonumber \\
& & - {i\over 2} (m - \widetilde{k} \cdot \widetilde{\gamma}) \gamma^- 
\int_0^{x^+} du e^{-ieA_-(u)(v+L)} {\cal E}(u,x^+;k^+,\widetilde{k}) 
\widetilde{\psi}_-(u,-L,\widetilde{k}) \Biggr\} , \qquad
\end{eqnarray}
where the higher dimensional mode function is,
\begin{equation}
{\cal E}(u,x^+;k^+,\widetilde{k}) \equiv \exp\Biggl[- {i\over 2} (m^2 +
\widetilde{k} \cdot \widetilde{k}) \int_u^{x^+} {{du'} \over {k^+ - eA_-(u') +
i/L}}\Biggr] \; .
\end{equation}
As for $D=2$, $\psi_-$ follows trivially from the Dirac equation,
\begin{equation}
\widetilde{\psi}_-(x^+,x^-,\widetilde{k}) = \Biggl({m - \widetilde{k} \cdot
\widetilde{\gamma} \over m^2 + \widetilde{k} \cdot \widetilde{k}} \Biggr)
\gamma^+ i \partial_+ \widetilde{\psi}_+(x^+,x^-,\widetilde{k}) \; .
\end{equation}

Of course we have learned nothing new about $QED_2$, which has long served as a
theoretical laboratory to model quark shielding and quark confinement 
\cite{cjs,coleman}. Indeed, it might be thought that our results conflict with
the received wisdom on this subject since we see pair creation for an electric
field of {\it any} strength, cf. our expression (\ref{probability}). It is 
quite well understood that this cannot be so. A clever energy argument due to 
Coleman \cite{coleman} shows that the addition of an $e^+ e^-$ pair at 
separation $R$ changes the electric field energy by,
\begin{equation}
{\Delta E} = \frac12 R \Bigl[ (E \pm e)^2 - E^2 \Bigr] \; . \label{sid}
\end{equation}
This is positive for $\vert E \vert < \vert e \vert/2$, so pair production is
not energetically favorable for a sufficiently weak field in $1+1$ dimensions. 

A little thought reveals that there is really no conflict with our results. 
Coleman's argument is based on including the electric fields of the produced 
pair. They give the term of order $e^2$ in (\ref{sid}). There is no doubt that
this is the right thing to do, but there is also no doubt that it is a higher
order effect. We worked at one loop, and the effect at that order is just the
particles' interaction with the background. This is the term of order $e E$ in
(\ref{sid}). If it were correct to retain only this term then it would be
energetically favorable to pull pairs out of the vacuum. So we are seeing what 
we should see in the one loop approximation: the imposition of a homogeneous 
electric field causes particle production, no matter how weak the field. It 
might be interesting --- and seems entirely within our reach --- to study how 
higher order effects conspire to stabilize the vacuum for sufficiently weak
electric background fields.

In higher dimensions the vacuum is unstable against pair production for any
nonzero value of $E$. This poses an interesting problem of back-reaction in
which the current of produced pairs initially reduces the electric field. What
should eventually happen is that a plasma forms and begins executing 
oscillations. This had to be studied numerically with previous treatments 
\cite{Kluger} because explicit expressions for the mode functions could not be 
obtained for a class of backgrounds wide enough to include the actual solution. 
The wonderful analytic control we have {\it for any} $E(x^+)$ ought to 
facilitate a less heavily numerical analysis. One would evaluate the 
expectation value of $J^-$, as a functional of $A_-$ and then use this as a
source in the relevant Maxwell equation,
\begin{equation}
-A_-^{\prime\prime}(x^+) = J^-[A_-](x^+,x^-) \; .
\end{equation}
However, our result (\ref{J-final}) is not immediately suitable for the task
because the large $L$ limit was taken at fixed background. Since this limit 
has $J^-$ diverge linearly in $L$, it must follow that the actual evolution of 
$A_-(x^+)$ depends upon $L$. This dependence will affect the limit used to 
compute $J^-$. It seems possible to untangle this problem, but we have not yet 
done so.

Of course our result (\ref{J-final}) for $J^-$ also depends upon $x^-$, and it
might be thought that this spoils the problem's homogeneity. In fact this is
not so because the $x^-$ dependence is restricted to an overall factor of $(L
+ x^-)$, and $L$ is going to infinity. Therefore, back-reaction becomes 
infinitely strong, infinitely fast in the large $L$ limit, and we can forget 
about the $x^-$ dependence.

It remains to comment on the curious fact that our results for the particle 
production probability (\ref{probability}), and for the two currents 
(\ref{J+final}-\ref{J-final}), possess an essential singularity at zero
electric field. This arose from taking the large $L$ limit. At finite $L$ the
expressions cannot be reduced to elementary functions, but they depend 
analytically upon the background field. It should also be noted that $J^-$
exhibits a new linear divergence at infinite $L$, derived from the pulse of
electrons created with uniform amplitude from all along the past $x^-$ axis. It
is tempting to regard these features as signals for an infinite volume phase 
transition in massive QED with a nonzero electric field. 

The background field formalism also gives one pause. Consider the expectation 
value of $J^+$ (\ref{J+final}). This could be represented, diagrammatically, as 
a single photon attached to a closed electron loop. We have been brought up to 
believe that differentiating such a diagram with respect to the background 
field attaches another photon. We have also been brought up to believe that 
photons couple to electrons with strength $e$. However, the actual computation 
reveals two terms,
\begin{eqnarray}
\lefteqn{{\delta \langle J^+ \rangle(x^+)  \over \delta A_-(y^+)} = - {e^2
\over 2 \pi} \delta(x^+ - y^+) e^{-2\pi \lambda(eA_-(y^+))} } \nonumber \\
& & \hspace{1cm} + {\partial \over \partial y^+} \Biggl\{ e^2 \lambda(e A_-(
y^+)) e^{-2\pi \lambda(e A_-(y^+))} \theta(y^+) \theta(x^+ -y^+) \Biggr\} 
\; . \quad
\end{eqnarray}
The first of these seems to represent an extra photon, but the second term 
gives {\it something} that seems to couple with strength $1$, and something 
{\it else} that couples with strength $1/e$. (Recall from (\ref{lambda}) that 
$\lambda = m^2/(2 \vert e \vert E)$.) We do not yet know what to make of this,
although it is certainly not an artifact of $1+1$ dimensions since the $3+1$
dimensional $J^+$ shows a very similar essential singularity \cite{ttw}.

\vspace{1cm}

\centerline{\bf Acknowledgements}
 
We have benefited from discussions with H. M. Fried, J. Iliopoulos, H. B.
Nielsen, M. Soussa and C. B.  Thorn. This work was partially supported by 
European Union grants HPRN-CT-2000-00122 and -00131, by the Greek General 
Secretariat of Research and Technology grant 97E$\Lambda$-120, by the DOE 
contract DE-FG02-97ER\-41029 and by the Institute for Fundamental Theory at 
the University of Florida. The authors express their gratitude for hospitality 
during recent visits to the CERN Theory Division, to the Laboratoire de 
Physique Th\'eorique of the Ecole Normale Superieure, and to the Department 
of Physics at the University of Crete.

\end{document}